\documentstyle[12pt,aaspp4]{article}

\def\nop{\noindent} 

\def\L{$\Lambda$}

\begin{document}

\title{Contributions to the Power Spectrum of Cosmic Microwave  \\
             Background from Fluctuations Caused by             \\
                    Clusters of Galaxies}

\author{S. M. Molnar\altaffilmark{1,2}}
\affil{Laboratory for High Energy Astrophysics, Code 662 \\
                   Goddard Space Flight Center \\
                      Greenbelt, MD 20771}

\author{M. Birkinshaw\altaffilmark{3}}
\affil{Department of Physics, University of Bristol, \\
    Tyndall Avenue, Bristol, BS8 1TL, UK}
 
\altaffiltext{1}{NAS/NRC Research Associate}

\altaffiltext{2}{previous address: Department of Physics, University of Bristol,
                                   Tyndall Avenue, Bristol, BS8 1TL, UK}

\altaffiltext{3}{also: Center for Astrophysics,
    60 Garden Street, Cambridge, MA 02138, USA}

\begin{abstract}

\nop We estimate the contributions to the cosmic microwave background radiation (CMBR)
power spectrum from the static and kinematic Sunyaev-Zel'dovich (SZ) effects, and from 
the moving cluster of galaxies (MCG) effect. 
We conclude, in agreement with other studies, that at sufficiently small scales
secondary fluctuations caused by clusters provide important contributions to the CMBR. 
At $\ell \gtrsim 3000$, these secondary fluctuations become important relative to
lensed primordial fluctuations.
Gravitational lensing at small angular scales has been proposed as a way to break the 
``geometric degeneracy'' in determining fundamental cosmological parameters. 
We show that this method requires the separation of the static SZ effect, 
but the kinematic SZ effect and the MCG effect are less important.
The power spectrum of secondary fluctuations caused by clusters of galaxies, 
if separated from the spectrum of lensed primordial fluctuations, might provide an
independent constraint on several important cosmological parameters.

\end{abstract}

Subject headings: 
cosmic microwave background --- 
galaxies: clusters: general --- 
methods: numerical ---

\section{Introduction}
\label{S:INTRO}

The power spectrum of the cosmic microwave radiation (CMBR) carries much cosmological
information about primordial density fluctuations in the early Universe.
As photons leave the last scattering surface and travel across the Universe, however, 
these brightness fluctuations are modified by intervening structures, causing secondary
fluctuations, which we would expect to become more important on small angular scales. 

The power spectrum of the CMBR alone can be used to determine cosmological parameters.
Recently it has been shown, however, that a geometrical degeneracy effect
prevents some combinations of cosmological parameters from being disentangled by 
the power spectrum alone (\cite{BondEfTeg97}; \cite{EfstathBond98}; \cite{ZaldarrSperSel97}). 
The primordial density fluctuations and matter content determine the positions
and magnitudes of the Doppler peaks at the last scattering surface. These fluctuations
are transferred to apparent angular scales determined by their angular diameter distance. 
As a result, cold dark matter (CDM) models with the same primordial density fluctuations, 
matter content, and angular diameter distance can not be distinguished. 
The models are ``effectively degenerate'' in the sense that their power spectrum is degenerate 
for parameter determination on intermediate and small scales. 
Although the observed power spectrum also depends on the time variation of the metric, via the 
integrated Sachs-Wolfe effect, this breaks the degeneracy only at large angular scales.
Unfortunately observations of the power spectrum do not provide strong constraints on models at 
large scales, since this is where the statistics of the data are dominated by the cosmic 
variance due to the fact that we have only one realization of our cosmological model, 
the Universe itself, and we encounter a sampling problem). 
Therefore we can determine, for example, only combinations such as $\Omega_0 h^2$ and 
$\Omega_b h^2$ (where $\Omega_0$ and $\Omega_b$ are the $z = 0$ matter and baryon density 
parameters and $h$ is the dimensionless Hubble parameter).

It has been noticed that gravitational lensing can break the geometric degeneracy at 
small angular scales, $\ell \gtrsim 2000$, in such a way that the cosmological parameters can be 
determined separately (\cite{MetcalfSilk98}; \cite{StomporEfstath98}). 
The effect of gravitational lensing on the CMBR was studied by several authors 
(see for example \cite{BlanchardSchneider87}; \cite{MetcalfSilk97}; \cite{Seljak97}).
Static gravitational lenses do not change a smooth CMBR, but the fluctuations get 
distorted by lensing. As a result, power from the acoustic peaks is transferred 
to small angular scales, conserving the variance of the spectrum. 
The amount of power transferred depends on the cosmological model, thus, in principle,
we can determine separately $\Omega_0$, $\Omega_b$, and $h$.
This method makes use of the small angular scale part of the power spectrum, 
where the amplitude of primordial fluctuations is declining and secondary fluctuations 
are becoming more important. The question naturally arises: 
How do contributions to the power spectrum from secondary fluctuations influence 
parameter determination based on the small scale CMBR power spectrum?

The most important secondary fluctuations are the thermal static and kinematic Sunyaev-Zel'dovich 
(SSZ and KSZ) effects (\cite{sz80}), the Rees-Sciama (RS) effect (\cite{ReesSciama68}), 
the moving cluster of galaxies (MCG) effect (\cite{BirkGull83}; \cite{GurvitsMitrof86}
; \cite{PyneBirkinshaw93}), point sources (\cite{ToffZAB99}), and, 
if the Universe was re-ionized at some early stage, the Ostriker-Vishniac effect
(\cite{OstrVish86}; \cite{Vish87}). 
In this paper we concentrate on secondary effects caused by clusters of galaxies.
Since detailed reviews are available on the SZ and the MCG effects 
(\cite{Reph95}; \cite{Birk98}), here we just summarize their major features. 
The thermal SZ effect is a change in the CMBR via inverse Compton scattering 
by electrons in the hot atmosphere of an intervening cluster of galaxies. 
We use the terms kinematic or static SZ effect depending on whether or not the intracluster gas
possesses bulk motion. 
To date only the static thermal SZ effect has been detected (\cite{Birk98}). 

The MCG effect is a special type of RS effect, due to the time-varying gravitational field of 
a cluster of galaxies as it moves relative to the rest frame of the CMBR. 
Unlike the original RS effect, the MCG effect is not caused by 
intrinsic variation of the gravitational field, so that in the rest frame of the 
cluster, the photons fall into and climb out of the same gravitational field. 
However, in the rest frame of the cluster the CMBR is not
isotropic, but has a dipole pattern, being brighter in the direction of the cluster
peculiar velocity vector. Photons passing the cluster are deflected towards its center.
Thus in the direction of the cluster peculiar velocity vector (ahead of the cluster)
one can see a cooler part of the dipole pattern. Towards the tail of the cluster, one can see a
brighter part of the dipole (ahead of the cluster). When transferring back to the
rest frame of the CMBR, we transfer the dipole out, but the fluctuations remain,
showing a bipolar pattern of positive and negative peaks. 
At cluster center there is no deflection, thus there is no effect.
The amplitude of the MCG effect is proportional to the product of the gravitational deflection 
angle and the peculiar velocity of the cluster.

The most important characteristics of the SSZ, KSZ, and MCG effects in the context of cosmology 
is that their amplitudes do not depend on the redshift of the clusters causing the effect.
Using thermodynamic temperature units, their maximum amplitude are about 500 $\mu$K, 20 $\mu$K, 
and 10 $\mu$K, respectively. The SSZ and KSZ effects have the same spatial dependence as the line 
of sight optical depth, the MCG effect has a unique bipolar pattern.
Assuming a King approximation for the total mass and an isothermal beta model for the intracluster
gas, the full width at half maximum (FWHM) of the SSZ and KSZ effects
$\approx (2 - 4)\,r_c$, where $r_c$ is the core radius, depending on $\beta$, 
which is typically between 2/3 and 1.
The MCG effect has a much larger spatial extent, with FWHM for each part of the bipolar 
distribution $\approx 10\, r_c$. 

The spectra of the effects are also important: the SSZ effect has a unique spectrum which 
changes sign from negative to positive at about 218 GHz.
The KSZ and MCG effects have the same frequency dependence as the primordial
fluctuations. The most important difference between the SZ effect and the MCG effect is that
the SZ effect is caused by intracluster gas, the MCG effect is caused by gravitational lensing
by the total mass regardless the physical nature of that mass. 
Therefore the SZ effect only arises from clusters with intracluster gas.
Clusters can produce significant MCG effects even if devoid of intracluster gas.

The effects of clusters of galaxies on the CMBR in a given cosmology have been a subject of 
intensive research since the late 1980s.  
There are several different ways of extracting information from these effects. 
Source counts of the SSZ effect were estimated by using the Press-Schechter mass function (PSMF) 
and scaling relations 
(\cite{ColeKaiser88}; \cite{Markevitchet92}; 1994; \cite{MakiSuto93}; \cite{BartSilk94}, 
De Luca, Desert \& Puget 1995, \cite{ColaMRV94}; 1997, \cite{sutoet99}). 
The importance of the SSZ effect was demonstrated and it was shown that thousands of detections 
are expected with the next generation of satellites. 
Contributions to the CMBR from the RS and the KSZ effects were derived by 
Tuluie, Laguna \& Anninos (1996) and Seljak (1996) for CDM models with zero cosmological constant. 
Tuluie et al. used N-body simulations and a ray-tracing technique, Seljak used N-body simulations 
and second order perturbation theory. Contributions from the SSZ and KSZ effects originating from 
large scale mass concentrations (superclusters) were studied by Persi et al. (1995).
Bersanelli et al. (1996), in their extensive study of the CMBR for the Planck mission, 
estimated the contribution to the power spectrum from the SSZ and KSZ effects. 
Aghanim et al. (1998) estimated the effects of the KSZ and MCG effects on the CMBR including
their contributions to the CMBR power spectrum.
Aghanim et al. simulated $12.5^\circ \times 12.5^\circ$ maps with pixel size of 
$1.5^\prime \times 1.5^\prime$. They used the PSMF normalized to X-ray 
data (assuming an X-ray luminosity-mass relation). The total mass was assumed 
to have a Navarro-Frenk-White profile (\cite{Navarroet97}), and the intracluster gas
was assumed to follow an isothermal beta model distribution. 
The time evolution of the electron temperature and the core radius were assumed to follow 
models of Bartlett \& Silk (1994), which are based on self-similar models of Kaiser (1986).
According to Aghanim et al. (1998)'s results, the KSZ effect is many orders of magnitude
stronger than the primordial CMBR on small angular scales, and therefore the effect would prevent
the use of the power spectrum to break the geometric degeneracy. 
Atrio-Barandela \& Mucket (1998) estimated the power spectra of the SSZ effect in 
a standard dark matter dominated model with different lower mass cut-offs.
Contributions to the power from the Ostriker-Vishniac effect in CDM models were estimated
by Jaffe \& Kamionkowski (1998).

In this paper we estimate the contributions to the CMBR power spectrum from the 
SSZ, KSZ, and MCG effects on small angular scales adopting cold dark matter dominated models. 
In our models we assumed scale invariant primordial fluctuations with
a processed spectrum having a power law form on cluster scales with a power law index 
of $n_P = -1.4$ (Bahcall \& Fan 1998). This maybe used as a first approximation as long as
contributions from very low and/or very high mass clusters are small
(cf. our discussion about mass cut offs at section~\ref{s:PSMF}).
We use three representative models in our study:
{\it Model 1}, open CDM (OCDM) model: 
a low density open model with $\Omega_0 = 0.2$, $\Lambda = 0$, $\sigma_8 = 1.2$;
{\it Model 2}, flat lambda CDM (\L CDM) model: 
a low density flat model with $\Omega_0 = 0.2$, $\Lambda = 0.8$, $\sigma_8 = 1.35$;
{\it Model 3}, standard CDM (SCDM) model:
a flat model with $\Omega_0 = 1$, $\Lambda = 0$, and $\sigma_8 = 0.65$.

In Section~\ref{s:method} we outline our method of estimating the power spectra of
secondary fluctuations caused by clusters of galaxies and discuss our normalization
method for the PSMF.
Sections~\ref{s:PSMF} and ~\ref{s:physpar} describe how we used the PSMF and the
scaling relations to obtain masses and other physical parameters of clusters.
In section~\ref{s:PowerSpectr} we present the spherical harmonic expansion of the SSZ, 
KSZ and MCG effects, and our method of estimating their power spectra.
Section~\ref{s:Simulation} describes our simulations to evaluate the integrals over clusters.
Sections~\ref{s:Results} and \ref{s:Discussion} present our results and discuss
the differences from previous work.

\section{Outline of the method}
\label{s:method}

\begin{figure} 
\centerline{
\plotone{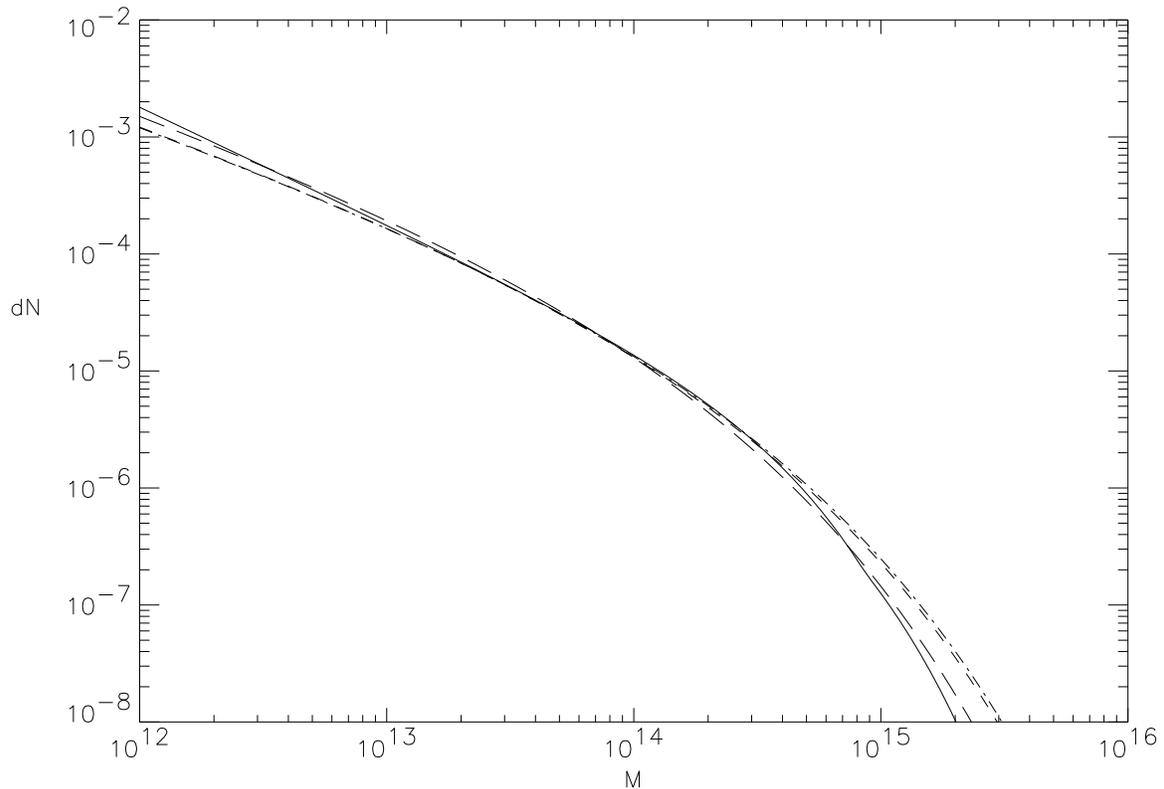}
}
\caption{
    The PSMFs, $dN/(dV d \ln M)(>M)$, at $z = 0$,
    used in our simulations compared to observations. 
    The long dashed, dash dot, and short dashed lines
    refer to the PSMFs of our $\Lambda = 0$, $\Omega_0 = 1$ (SCDM); 
    $\Lambda = 0$, $\Omega_0 = 0.2$ (OCDM); 
    and $\Lambda = 0.8$, $\Omega_0 = 0.2$ (\L CDM) models respectively. 
    The solid line represents the observationally-derived mass function 
    of Bahcall and Cen (1993).
\label{F:PSMF_OBS} 
} 
\end{figure}

\begin{figure} 
\centerline{
\plotone{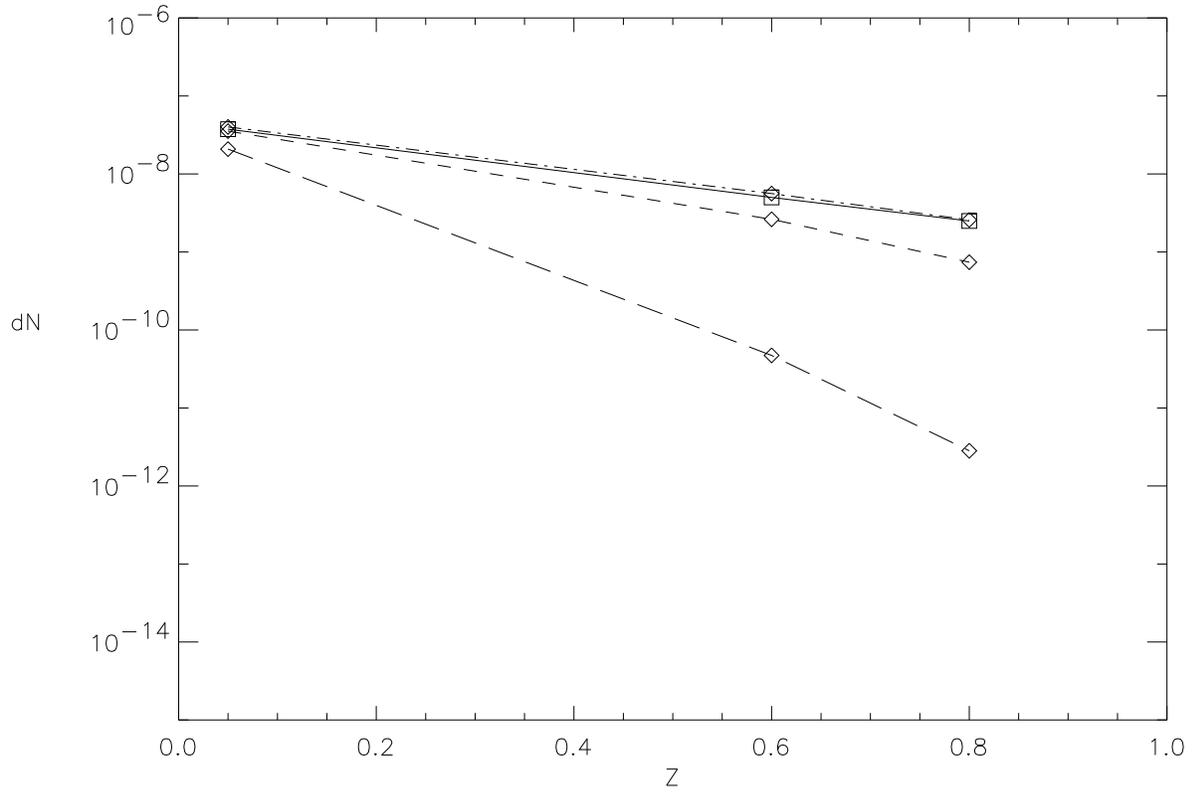}
}
\caption{
    High mass ($M > 1.6 \times 10^{15} M_\odot$) PSMFs, 
    $d N/ (dV dz)$ in units of number Mpc$^{-3}$, as a function of redshift, $z$, 
    used in our simulations compared to observations.
    The long dashed, dash dot, and short dashed lines
    refer to the cumulative PSMFs of our $\Lambda = 0$, $\Omega_0 = 1$ (SCDM); 
    $\Lambda = 0$, $\Omega_0 = 0.2$ (OCDM); and $\Lambda = 0.8$, $\Omega_0 = 0.2$ 
    (\L CDM) models respectively.
    The solid line represents the mass function derived from observations 
    (Bahcall et al. 1997), which is a close match to our OCDM model.
\label{F:PSMF_Z_OBS} 
} 
\end{figure}

We used the Press-Schechter Mass Function (PSMF, \cite{presschechter74}) as a distribution 
function for cluster masses. We used $n=-1.4$ as indicated by 
observations (Bahcall and Fan 1998, and references therein). 
We used observationally determined cluster abundances as a constraint on the PSMF. 
Where necessary, we altered the model parameters resulting from the usual top hat spherical 
collapse model since that model is only an approximation. 
In our SCDM model we changed only the overall normalization of the PSMF by multiplying it by 0.23 
(a similar normalization was used by \cite{DeLucaet95}). This procedure is inconsistent with the
interpretation of the PSMF as a probability distribution (it does not integrate to unity),
but we use results for the SCDM model only as a comparison to the other two models. 
In our Lambda-CDM model we multiplied the critical density threshold, $\delta_c^0$, 
obtained from the spherical collapse model (equation~[\ref{E:delta_c}]), by 1.23 
(which is equivalent to changing the $\sigma_8$ normalization) and made no other changes. 
Our OCDM model needed no adjustments.
With these changes all three models agree well with the present day ($z=0$) observed mass spectrum 
(Figure~\ref{F:PSMF_OBS}; Bahcall \& Cen 1993). 
For high masses, the first two models (OCDM, and \L CDM) also agree with the observed $z$ dependence 
of the high mass cumulative mass function (Figure~\ref{F:PSMF_Z_OBS}; Bahcall, Fan \& Cen 1997).
The \L CDM and the SCDM models agree with CMBR constraints, while 
the OCDM model is rejected by these constraints (\cite{LinewBar98}).
As a cautionary note, it is useful to keep it in mind that taking {\it all} data
into account none of these models are acceptable. 

We assumed that the total mass distribution follows a truncated King profile 
(\cite{King62}). For the intra-cluster gas we assumed an isothermal beta model
(Cavaliere \& Fusco-Femiano 1976). Isothermal beta model fits to X-ray images of clusters give
$\beta \approx 2/3$ (\cite{JonesForman84}). Determinations of the $\beta$ parameter
based on spectroscopy suggest $\beta \approx 1$ (\cite{Girardiet96}; \cite{LubinBahcall93}). 
Numerical simulations imply a range for $\beta$ from 1 to about 1.3
($\beta = 1.05$: \cite{EvrardMetNav96}); $\beta \approx 1.3$: \cite{BryanNorman98}; 
\cite{Frenket99} suggest $\beta = 1.17$). 
We follow the precepts of \cite{Ekeet98} and adopt $\beta = 1$. 
This choice provides a mass temperature function which is in a good
agreement with the observed function (\cite{Horneret99}). The fitted
X-ray spatial distribution is highly dependent on the X-ray structure
of the core, and may be expected to be less reliable in the outer
regions of clusters. The SZ effect is more sensitive to the outer
regions (the SZ effect is proportional to the electron density as
opposed to thermal bremsstrahlung, which is proportional to density
squared). Choosing the spectroscopically derived $\beta = 1$ gives a
smaller SZ effect, and so our choice of $\beta$ should provide a
conservative estimate of the contribution of the SZ effect to the
power spectrum. A slightly larger $\beta$ would not change our results
significantly. Although the beta model describes the inner
intra-cluster gas well, for more accurate SZ work an improved cluster
model, which fits the outer regions better, will be needed.

The other physical parameters were determined using the virial theorem, a
spherical collapse model, and models of the intra-cluster gas by Colafrancesco \& Vittorio (1994).
We assumed a Maxwellian distribution for the peculiar velocities, $v_{pec}$, and 
used results of N-body simulations by Gramann et al. (1995) to normalize the distribution.
We took velocity bias into account, and assumed that the peculiar velocities 
are isotropically distributed. 

We used analytical approximations to calculate the contributions of the SSZ, KSZ and MCG effects
to the CMBR power spectrum. 
These contributions are important only at small angular scales, where we can neglect the overlap 
between cluster images and ignore the weak cluster-cluster correlation, 
and therefore we can approximate the resulting power spectrum by summing
the contributions from individual clusters (similar methods were used by 
\cite{ColeKaiser88}; \cite{BartSilk94}; and \cite{AtBaMuck99}).
We expanded the SSZ, KSZ, and MCG effects as Laplace series (i.e., series of spherical harmonics), 
then determined the individual cluster contributions and summed over the clusters
(for a detailed description see \cite{Molnar98}).

Our approximation breaks down at large angular scales, but at these scales primordial
fluctuations dominate the CMBR, and the cluster contribution is only a minor perturbation,
so that only a rough indication of the cluster effect is needed.

\section{Total Mass Distribution}
\label{s:PSMF}

According to the Press-Schechter method, the co-moving number density of clusters of 
total mass $M$ at redshift $z$ (the PSMF) is given by

\begin{equation} \label{E:PSMF}
    { d\, n_c(M, z) \over  d\, \ln M} = \sqrt{ 2 \over \pi } \, { \Omega_0 \rho_c \over M } \, 
                           \nu(M, z)  \,
                          \Bigl(- { d \, \ln \sigma \over d\, \ln M} \Bigr) \, e^{-\nu(M,z)^2/2}
,
\end{equation}
where $\Omega_0$ is the matter density today in units of the critical density, 
$\rho_c = 1.88 \times 10^{-29} \, \rm g\, cm^{-3}$ is the current critical density of the universe
(we adopt a dimensionless Hubble parameter $h_{100} = 0.5$ in our work); 
$\nu(M, z)  = \delta_c^0(\Omega_0, z) /\sigma(M)$, 
where the present mass variance for a power law power spectrum, $P(k) \propto k^n$, is

\begin{equation}
   \sigma(M) = \sigma_{8} \,  \Biggl( { M \over M_8} \Biggr)^{-\alpha}
,
\end{equation}
$\alpha = (n + 3)/6$, $M_8 = 6 \times 10^{14} \Omega_0\, h^{-1}M_\odot$ 
is the mass within an $R_8 = 8 h^{-1}$ Mpc sphere, and $\sigma_8$  is the normalization
(Lacey \& Cole 1993, Press \& Schechter 1974).
The over-density threshold linearly extrapolated to the present may be expressed as
(\cite{LaceyCole93}; \cite{Navarroet97})

\begin{equation} \label{E:delta_c}
    \delta_c^0(\Omega_0, z) = \cases{
                      { 3 \over 20} (12 \pi )^{2/3} ( 1 + z) & $\Omega_0 = 1$, $\Lambda = 0$ \cr
                      {3 \over 2} D(\Omega_0, 0) \Bigl[ 
                     \bigl( { 2 \pi \over \sinh \eta -  \eta } \bigr)^{2/3} + 1 \Bigr]       & 
                                                                     $\Omega_0 < 1$, $\Lambda = 0$ \cr
                       0.15 (12 \pi)^{2/3} \Omega_m^{0.0055}
                       D_\Lambda(\Omega_0, 0)/D_\Lambda(\Omega_0, z)     & 
                                                      $\Omega_0 < 1$, $\Lambda = 1 -  \Omega_0$ \cr 
                              } 
,
\end{equation}
where the conformal time for open models is

\begin{equation}
 \eta = \cosh^{-1} 
           \Biggl[ {2 \over  \Omega_m(z) } - 1 \Biggr]
.
\end{equation}
For open models with cosmological constant $\Lambda = 0$ 
the linear growth factor is (Peebles 1980)

\begin{equation} \label{E:Dz_LAM0}
   D(\Omega_0, z) = 1 + { 3 \over w} + { 3 (1 + w)^{1/2} \over w^{3/2}}
                                 \ln \Bigl( (1 + w)^{1/2} - w^{1/2} \Bigr)
, 
\end{equation}
where $w(\Omega_0, z) = \Omega_m^{-1}(z) - 1$, 
and the density parameter $\Omega_m(z)$ is

\begin{equation}
   \Omega_m (z) = 
         {\Omega_0 (1 + z) \over \Omega_0 (1 + z) + (1 - \Omega_0) } 
.
\end{equation}
For models with a non-zero cosmological constant
the integral can not be done analytically, thus we use
an approximation (Lahav et al 1991; Caroll, Press \& Turner 1992) to obtain
the equivalent expression to (\ref{E:Dz_LAM0}), 

\begin{equation}
   D_\Lambda(\Omega_0, z) = ( 1 + z )^{-1} 
        { 5 \Omega_f(z) \over 2}
                \Biggl\{  \Omega_f(z)^{4/7} - \Omega_\Lambda(z) + 
                           \Biggl[ 1 + {\Omega_f(z)                   \over 2} \Biggr]
                           \Biggl[ 1 + {\Omega_\Lambda(z) \over 70} \Biggr] \Biggr\}^{-1}
.
\end{equation}
The density parameters, $\Omega_f (z)$ and $\Omega_\Lambda (z)$, 
for spatially flat Universes with $\Lambda = 1 -  \Omega_0$, are 

\begin{eqnarray} 
  \Omega_f (z)             &  = {\Omega_0 (1 + z)^3 \over \Omega_0 (1 + z)^3 + 1 - \Omega_0 } \\
  \Omega_\Lambda (z) &  = {1 - \Omega_0  \over \Omega_0 (1 + z)^3 + 1 - \Omega_0 } 
.
\end{eqnarray}
The normalization of these growth functions is chosen so that at high redshifts they approximately
match the time variation of density contrast in an Einstein-de Sitter ($\Omega_0 = 1$) 
Universe, which is a good approximation to the early Universe whatever its density parameter today.
The total number of clusters at redshift $z$ (in a redshift interval of $dz$) is

\begin{equation}
  N(z) dz = \int_{M_{low}}^{M_{up}} {d n_c(M, z) \over d M} \; { dV \over dz} \; d M d z
,
\end{equation}
where $M_{low}$ and $M_{up}$ are the lower and upper mass cut offs for clusters.
We used $M_{low} = 10^{13} \; M_\odot$ for the SSZ and KSZ effects, 
$M_{low} = 10^{12} \; M_\odot$ for the MCG effect, and 
$M_{up} = 1 \times 10^{16} \; M_\odot$ for all effects.
The lower cut off, $M_{low}$, for SZ effects signifies the lowest cluster mass for which 
we expect a well-developed intracluster atmosphere. In the case of the MCG effect, $M_{low}$ 
is the mass limit from which we consider a mass concentration as a cluster (``formation''). 
We found that low mass clusters do not contribute substantially to the power spectrum, 
thus the lower cut off, $M_{low}$, has little effect on our results. 
The upper cut off, $M_{up}$, has no effect on our results (as long as it is large enough): 
the probability of getting such a large cluster is negligible, so that the contribution 
from more massive clusters is negligible.

\section{Other physical parameters of clusters of galaxies}
\label{s:physpar}

We assumed a truncated King profile for the total mass distribution

\begin{equation} \label{E:King_ro}
  \rho(r) = \cases{ 
                    \rho_0 \biggl( 1+ {r^2 \over r_c^2} \biggr)^{-{3\over 2}} & $ r <  p r_c$\cr
                                   0                                           & $ r \ge p r_c$}
\end{equation}
where $\rm r_c$ is the core radius, $p r_c$ is the cut off, 
and an isothermal $\beta$ model for the intra-cluster gas

\begin{equation} \label{E:n_el}
    n(r) = \cases{
               n_0  \biggl(1+{r^2 \over r_c^2}\biggr)^{-{3\beta \over 2}} & $ r < p r_c$      \cr
                          0                                             & $ r \ge p r_c$     }
,
\end{equation}
where $n(r)$ and $n_0$ are the electron number density at radius $r$ and at the
center of the cluster (Cavaliere \& Fusco-Femiano 1976).
Analytical studies and numerical simulations show that the gas density profile scales
with the total density, and that the gas central electron density may be expressed as

\begin{equation} \label{E:n_el_0}
  n_0 = f_g  {  2 \rho_0 \over m_p (1 + X) }
,
\end{equation}
where $X=0.69$ is the average Hydrogen mass fraction, $f_g$ is the intra-cluster gas
mass fraction. $\rho_0$, the central mass density, is determined from the total 
mass by integrating equation~(\ref{E:King_ro}).
Little is known about the total mass and redshift dependence of the intra-cluster gas from 
observations. Here, we adopt Colafrancesco \& Vittorio (1994)'s model which assumes that changes 
in the intra-cluster gas are driven by entropy variation and/or shock compression and heating.
According to their model, the gas mass fraction may be expressed as

\begin{equation}
  f_g = f_{g0} \Biggl( { M \over  10^{15} h^{-1} M_\odot } \Biggr)^\eta ( 1 + z)^{-s}
,
\end{equation}
where the normalization, $f_{g0} = 0.1$, is based on local rich clusters, and we
used $\eta = 0.5$ and $s = 1$, which are consistent with available data.
Using the virial radius to express the core radius, $r_c = R_v/p$, and assuming 
spherical collapse, we obtain

\begin{equation} \label{E:r_cor}
  r_c(\Omega_0, M, z) = {1.69 h^{-1} {\rm Mpc} \over p} 
            \Biggl[  \Biggl( {M \over 10^{15} h^{-1} M_\odot } \Biggr)
                        { 178  \over \Omega_0 \Delta_c(\Omega_0, z) } \Biggr]^{1/3} 
                          {1 \over 1 + z } 
,
\end{equation}
where $\Delta_c(z) \equiv \rho_v(z) / \rho_b(z)$ 
is the overdensity of the cluster relative to the background (\cite{ColaBlasi98}). 
For $\Lambda = 0$ models (our OCDM and SCDM models) the over density may be expressed as

\begin{equation} \label{E:Delta_c_L0}
     \Delta_c(\Omega, z) = 4 \pi^2 Q^2 
                    \Bigl[ ( Q^2 + 2Q)^{1/2} - \ln \bigl(1 + Q + ( Q^2 + 2Q)^{1/2} \bigr) \Bigr]
,
\end{equation}
where $Q = 2( 1 - \Omega_0)/ (\Omega_0 (1+z))$ (Oukbir \& Blanchard 1997).
For spatially flat models with finite cosmological constant (our \L CDM model) we have

\begin{equation} \label{E:Delta_c_L}
     \Delta_c(\Omega, z) = 18 \pi^2 \biggl[1 + 0.4093 
                                           \bigl( \Omega_f(z)^{-1} - 1 \bigr)^{0.9052} \biggr]
,
\end{equation}
where we used the approximation of Kitayama \& Suto (1996).

Numerical models of cluster formation show that cluster temperature scales
with total mass. Using the virial theorem and assuming spherical collapse with
a recent-formation approximation in a standard CDM model, the electron temperature, 
$T_e$, becomes

\begin{equation} \label{E:T_eOM1}
  k T_e = 7.76 \, \beta^{-1} 
          \Biggl( { M \over 10^{15} h^{-1} M_\odot } \Biggr)^{2/3} \, ( 1 +z )\, {\rm keV}
,
\end{equation}
where $\Delta_c$ is the density contrast of a spherical top-hat perturbation 
relative to the background density just after virialization 
(cf. for example Eke, Cole \& Frenk 1996).
The recent-formation approximation, however is valid only for $\Omega_0 = 1$.
For our low matter density open model ({\it Model 1}, OCDM), 
we use a model which takes into account accretion during the evolution of clusters, 
and leads to the following scaling:

\begin{equation} \label{E:T_eOPEN}
  k T_e = 2.76 \, \beta^{-1} {1 - \Omega_0 \over \Omega_0^{2/3}}
         \Biggl( { M \over 10^{15} h^{-1} M_\odot } \Biggr)^{2/3}
         \Biggl[ \biggl( {2 \pi  \over \sinh \eta - \eta } \biggr)^{2/3} + 
                                           {n_P + 3 \over 5} \Biggr] \, {\rm keV}
.
\end{equation}
This was derived for open models with zero cosmological constant (\cite{VoitDonahue98}), 
but since structure formation evolves similarly in a low density model with the same matter density 
and a zero cosmological constant, we use it for our {\it Model 2} (\L CDM) as an approximation.

We assumed a Maxwellian for the cluster peculiar velocity distribution, $v_{pec}$, 
as expected from a Gaussian initial density field:

\begin{equation} \label{E:P_v_pec}
   P(v_{pec}, z)\,  d v_{pec} \propto
          v_{pec}^2 \, \exp \bigl\{ - v_{pec}^2/2 \sigma_p(z)^2 \bigr\} d v_{pec} 
,
\end{equation}
where $\sigma_p(z)$ is the Maxwellian width of the peculiar velocity distribution.
The rms peculiar velocity from linear theory, smoothed with a top-hat
window function of radius $R$, $W_R$, is given by 

\begin{equation} \label{e:pecvel1}
  \langle v^2 \rangle _R (z) = H^2(z)\,  a^2(z) \,  f^2(\Omega_0, \Lambda) \, \sigma_{-1}(R)
,
\end{equation} 
where $a(z)$ is the scale factor, and the moments, $\sigma_j(R)$, are defined as

\begin{equation}
   \sigma_j(R) =  { 1 \over 2 \pi^2 } \int_0^\infty k^{2j + 2} P(k) W(kR) dk
,
\end{equation}
where $P(k)$ is the Fourier transform of the power spectrum and 
equation~(\ref{e:pecvel1}) uses the moment $j = -1$ (\cite{Peebles80}). 
The velocity factor, $f(z) \equiv d \ln \delta / d \ln a$, can be approximated as 
(\cite{Peebles80}; 1984)

\begin{equation} 
        f(z) \approx \cases{
                       \Omega_m^{0.6} (z)  &  $\Lambda = 0$ \cr
     \Omega_f(z) \Bigl[ {5 \over 2 (1+z) D_\Lambda(\Omega_0, z)  } -  { 3 \over 2} \Bigr]  &  
                                                                      $\Lambda = 1 - \Omega_0$ \cr
                                                            } 
.
\end{equation}
The cluster peculiar velocity rms differs from this since we assume that clusters form at the 
peaks of the density distribution, and with this bias may be expressed as

\begin{equation}
   \langle v_p^2 \rangle_R (z) = \langle v^2 \rangle_R (z) 
                           \biggl[ 1 - { \sigma_0^4(R) \over \sigma_1^2(R) \sigma_{-1}^2(R) } \biggr]
\end{equation}
(Bardeen et al 1986). Colberg et al. (1998) calculated the velocity bias in a series of
CDM models using a top-hat filter and processed CDM power spectra. 
The correction factor has a weak dependence on $\Omega_0$: it is about 0.8 for low density 
and flat CDM models. 

We obtain the Maxwellian width in equation~(\ref{E:P_v_pec}) from the rms peculiar velocity 
from averaging a Maxwellian:

\begin{equation} \label{E:sigm_v_pec}
  \langle v_p^2 \rangle = { \int_0^\infty  v^4 \, \exp\{ - v_{pec}^2/2 \sigma_p(z)^2\} dv
                         \over 
                         \int_0^\infty  v^2 \, \exp\{ - v_{pec}^2/2 \sigma_p(z)^2 \} dv }
                   = 3 \sigma_p^2
.
\end{equation}
We expressed $\sigma_p$ as 
$\sigma_p = norm \times [ H(z) a(z) f(z)) ]/[ H(0) a(0) f(0) ]$.
The normalization at $z = 0$ was determined by using results on the peculiar velocity 
distribution from numerical simulations (\cite{Gramannet95}). 
Thus we obtain the following expression for the Maxwellian width of the peculiar velocities, 
$\sigma_p(\Omega_0, \Lambda, z)$, 
with velocity bias for models with no cosmological constant (OCDM, SCDM, $\Lambda = 0$):

\begin{equation} \label{E:sigma_p_L0}
  \sigma_p(\Omega_0, 0, z)  = (410 \, {\rm km\,s^{-1}} )\, \Omega_m^{0.6} (z) 
                          \bigl( \Omega_0 ( 1 + z) + 1 - \Omega_0 \bigr)^{1/2}
.
\end{equation}
For our \L CDM model ($\Lambda = 1 - \Omega_0$) we obtain

\begin{eqnarray}   \label{E:sigma_p_LAM}
  \sigma_p(\Omega_0, 1 - \Omega_0, z)  &  = & 410 \, {\rm km\,s^{-1}}\, {1 + z \over ( \Omega_0 (1+z)^2 + 1 - \Omega_0)^{1/2} } \, {D_\Lambda(\Omega_0, 0)  \over D_\Lambda(\Omega_0, z) } \, \times  \nonumber \\
 &      &   \nonumber \\
&\times & \Biggl[ {5 - 3 (1+z) D_\Lambda(\Omega_0, z) \over 5 - 3 D_\Lambda(\Omega_0, 0) } \Biggr], \\
 &      &   \nonumber
\end{eqnarray}
This normalization is significantly larger than some recent measurements suggest
(Bahcall \& Oh 1996), but it is a good match to others (Gramann 1998). 
This uncertainty should be remembered when interpreting our final results.

\section{Power Spectra of SSZ, KSZ and MCG Effects}
\label{s:PowerSpectr}

Ignoring the correlation between clusters, the power spectrum becomes

\begin{equation} \label{e:c_ell_int}
  C_\ell^X = \int dz \int dM \; {d n_c(M, z) \over d M}\; G_\ell^X\; { dV \over dz}
,
\end{equation}
where $G_\ell^X$ is the contribution from clusters with total mass $M$ at
redshift $z$, and $X$ denotes the SSZ, KSZ or MCG effects.
$dV/dz$ is the differential volume element (assumed isotropy)

\begin{equation} \label{E:dV_dz}
   { dV \over dz} = r(z)^2  { 4 \pi c \over H_0} \biggl[ \Omega_0 (1 + z)^3 +
             (1 - \Omega_0 - \Lambda) (1 + z)^2 + \Lambda \biggr]^{-1/2}
,
\end{equation}
where the effective distance $r(z)$ is 

\begin{equation} \label{E:r_z_dist}
   r(z) = \cases{ { 2 c \over H_0 } \Bigl[{ \Omega_0 z + (\Omega_0 -2) 
          (\sqrt {1 + \Omega_0 z} -1) \over \Omega_0^2 (1 + z)} \Bigr] 
          & $\Lambda = 0$\cr
          {c \over H_0 } \int_0^z dx \bigl[\Omega_0 (1 + x)^3 + 1 - \Omega_0 
          \bigr]^{-1/2}
          & $\Lambda =  1 - \Omega_0$}
\end{equation}
(Peebles 1993). 
In general, the coefficients $G_\ell$ may be determined by calculating the spherical 
harmonic expansion of the cluster image by averaging out the azimuthal parameter, $m$, 

\begin{equation}
  G_\ell = { 1 \over 2 \ell +1}  \sum_{m = -\ell}^\ell\, | a_{\ell m} |^2 
.
\end{equation}
Our task is to determine the $a_{\ell m}$ coefficients.

The SSZ and KSZ effects are cylindrically symmetric for spherical clusters, 
therefore we may describe them using only one coordinate, the angular distance from 
the cluster center. We separate the effects into amplitudes and geometrical form factors which 
carry their spatial dependence.
The SSZ and KSZ effects in thermodynamic temperature units may be
expressed as

\begin{eqnarray}
  \Biggl( {\Delta T \over T } \Biggr)_S(x, \theta)  \equiv  \Delta_S & =  & \Delta_S^0(x) \zeta(\theta) \\
  \Biggl( {\Delta T \over T }\Biggr)_K(x, \theta)  \equiv  \Delta_K  & =  & \Delta_K^0(x) \zeta(\theta) 
,
\end{eqnarray}
where the central effects for the SSZ and KSZ effects are

\begin{equation} \label{E:Delta_I_S_R}
  \Delta_S^0(x, \Theta) = \bigl[ Y_0(x) + \Theta Y_1(x) + \Theta^2 Y_2(x) + \Theta^3 Y_3(x) +
                         \Theta^4 Y_4(x)  \bigr] \, \Theta \, \tau_0
,
\end{equation}
and

\begin{eqnarray} \label{E:Delta_I_K_R}
     \Delta_K^0(x, \Theta) & =  & \Biggl\{
      \Bigl[ {1 \over 3} Y_0 + \Theta \Bigl( {5 \over 6} Y_0 + {2 \over 3} Y_1 \Bigr) \Bigr] \, \beta^2 
      - \bigl[1 + \Theta C_1(x) + \Theta^2 C_2(x) \bigr]\, \beta \, P_1(\alpha)                       \\
     & + &  \bigl[ D_0(x) + \Theta D_1(x) \bigr]  \, \beta^2 \, P_2(\alpha) \Biggr\}\, \tau_0  \nonumber
.
\end{eqnarray}
In these expressions $P_k$ is the Legendre polynomial of order of $k$, 
$x = h \nu / k_B T_{CB}$ is the dimensionless frequency,
$\Theta = k_B T_e/(m_e c^2)$ is the dimensionless temperature, 
$\alpha$ is the angle between the cluster's peculiar velocity vector and its position vector, 
$h$, $\nu$, $k_B$ and $T_{CB}$ are the Planck constant, frequency, Boltzmann constant,
and temperature of the CMBR, $T_{CB} = 2.728 \pm 0.002$ K (\cite{Fixsenet96}), 
and the lengthy expressions for the spectral functions $Y_i(x)$, $C_i(x)$ and $D_i(x)$
may be found in Nozawa, Itoh \& Kohyama (1998).
These functions arise from an expansion of the Boltzmann equation and although they are inaccurate
for high temperature clusters, their precision is sufficient 
for our purposes here (for a discussion see \cite{MolnarBirkinshaw99} and references therein).
The optical depth through the cluster center for gas model (\ref{E:n_el}) is

\begin{equation}
   \tau_0 = 2 \sigma_T n_0 r_c \, I_p(3 \beta/2, p)
,
\end{equation}
and the geometrical form factor is

\begin{equation}
   \zeta(\theta) =  \int (n_e/n_0) dz = \bigl( 1 + d_c^2 \bigr)^{ {1 \over 2} - {3 \beta \over 2} }
                            \,j(\beta, p, d_c)
,
\end{equation}
where the function $j(\beta, p, d_c)$ is defined as 

\begin{equation}
 j(\beta, p, d_c) = { I_p( 3 \beta/2,\sqrt{ p^2 - d_c^2}) \over I_p(3 \beta/2, p) }
,
\end{equation}
$d_c = \theta/\theta_c$ holds in the small angle approximation, and the integral, 
$I_p(\alpha, q)$, is 

\begin{equation}
  I_p (\alpha, q) = {\sqrt{\pi} \over 2} { \Gamma( \alpha - { 1 \over 2} ) \over \Gamma( \alpha )} - 
                    (q + 1)^{-\alpha} (q - 1)^{1-\alpha} 
                    {\Gamma( 2 \alpha- 1 ) \over \Gamma( 2 \alpha ) } F(\alpha; 1; 2 \alpha; 2 q)
,
\end{equation}
where $\Gamma$ is the gamma function, $F(x; y; w; z)$ is Gauss' hyper-geometric function, and 
$\alpha$ must be greater than 1/2 (\cite{GradshtRyzh80}).
The geometrical form factor is normalized to one at the cluster center ($\zeta(0) = 1$). 
We may expand the SSZ and KSZ effects in Legendre series as 

\begin{equation}
  \Delta_{SZ} = \Delta_{SZ}^0\, \sum_{\ell=0}^\infty { 2 \ell + 1 \over 4 \pi } \zeta_\ell P_\ell
,
\end{equation}
where $\zeta_\ell$ are Legendre coefficients of $\zeta$, and 
$SZ$ refers to the SSZ or the KSZ effect.
We determine the Legendre coefficients using a small angle approximation, as

\begin{equation} \label{E:zeta_ell}
   \zeta_\ell = 2 \pi \int_0^{p \theta_c} \zeta(\theta) 
                                   J_0[(\ell + {1\over 2}) \, \theta ] \, \theta \, d \theta 
,
\end{equation}
where we used the approximation

\begin{equation}
   P_\ell \approx J_0[(\ell + {1\over 2}) \, \theta ]
,
\end{equation}
where $J_0$ is a Bessel function of the first kind and zero order (\cite{Peebles80}).
We can convert Legendre coefficients to Laplace coefficients by expressing the
Laplace series of such a function as 

\begin{equation}
   \sum_\ell a_{\ell 0} Y_\ell^0(\theta, \varphi) = 
                              \sum_\ell a_{\ell 0} \sqrt { 2 \ell + 1 \over 4 \pi} P_\ell(\mu) = 
                         \sum_\ell { 2 \ell + 1 \over 4 \pi } \, b_\ell P_\ell(\mu)
,
\end{equation}
where $\mu = \cos \theta$, and $Y_\ell^m$ and $P_\ell$ are the 
spherical harmonics and Legendre polynomials.
Therefore the conversion can be done as

\begin{equation}
   a_{\ell 0} = \Bigl( { 2 \ell + 1 \over 4 \pi} \Bigr)^{1/2} b_\ell 
.
\end{equation}
Thus the Laplace series of the SSZ and KSZ effects  become

\begin{equation} \label{E:Delta_SZ}
 \Delta_{SZ} (\theta) = \Delta_{SZ}^0 \sum_\ell \Bigl( { 2 \ell + 1 \over 4 \pi} \Bigr)^{1/2} 
                                                               \zeta_\ell Y_\ell^0 (\theta) 
.
\end{equation}

Using equation~(\ref{E:Delta_SZ}), the contribution of one cluster to the power spectrum 
of the SSZ and KSZ effects becomes

\begin{equation}
  G_\ell^{SZ} = {1 \over 4 \pi } (\Delta_{SZ}^0)^2 \zeta_\ell^2
,
\end{equation}
where $\Delta_{SZ}^0$ and $\zeta_\ell$ are given by 
equations~(\ref{E:Delta_I_S_R}), (\ref{E:Delta_I_K_R}), and (\ref{E:zeta_ell}).

Similarly, the MCG effect may be expressed as

\begin{equation} \label{E:Delta_T_MCG}
   \Biggl( {\Delta T \over T } \Biggr)_M(x, \theta, \varphi ) \equiv  \Delta_M  = 
                                                      \Delta_M^{max} \xi (\theta,\varphi )
,
\end{equation}
where $\xi (\theta,\varphi )$ is the geometrical form factor. $\theta$ is the angle of the line 
of sight relative to the center of the cluster. The azimuthal angle, $\varphi$, 
is measured in the plane of the sky
from the direction of the tangential component of the peculiar velocity.
The maximum of the MCG effect is 

\begin{equation}
   \Delta_M^{max} = - (v_p/c) \sin \alpha\, \delta_{max}
,
\end{equation}
where $\delta_{max}$ is the maximum deflection angle, $\alpha$ is the angle between the 
cluster's peculiar velocity vector, $v_p$, and its position vector, 
and $c$ is the speed of light in vacuum.
For a spherically symmetric thin lens the deflection angle is given by 

\begin{equation}  \label{E:delta_MCG0}
  \delta(d) = { 4 G \over c^2 \, d } {\cal M}(d)
,
\end{equation}
where $d$ is the impact parameter at the source, 
and $\cal M$ is the mass enclosed by a cylindrical volume with axis parallel to the
line of sight and radius equal to the impact parameter $d$
(cf. for example Schneider, Ehlers \& Falco 1992).
Using the King approximation for the density distribution (equation~[\ref{E:King_ro}]),
the total mass in cylindrical coordinates, $(R, \psi, z)$, becomes

\begin{equation}
 {\cal M} =  \rho_0 r_c^3 \,  \int_0^{2 \pi} d \psi
                  \int_{-d_c}^{d_c} dz \int_0^p\, 
                  { R dR  \over \bigl(1 + R^2  + z^2 \bigr)^{3 \over 2} }
,
\end{equation}
where $d_c \equiv d/r_c \approx \theta/\theta_c$ in the small angle approximation.
A straightforward integration and equation~(\ref{E:delta_MCG0}) lead to

\begin{equation}
  \delta(d_c, p) = { 4 G {\cal M} \over c^2 r_c g(p, p) }\, { g(d_c, p) \over d_c }
,
\end{equation}
where the function $g(x, p)$ is 

\begin{equation}
   g(x, p) =  \ln ( 1 + x^2) + 
                        \ln \Biggl[ { p + \sqrt{ 1 + p^2} \over  p + \sqrt{ 1 + p^2 + x^2 } } \Biggr]
.
\end{equation}
Thus the geometrical form factor in our case becomes

\begin{equation} \label{E:xi_g}
   \xi(\theta, \varphi) = { g(d_c, p) \over b_m d_c} \cos \varphi 
,
\end{equation}
where $b_m$ is the maximum value of the function $g(x, p)/x$.

The MCG effect depends only on $\cos \varphi$, therefore we need to determine only the 
$m = \pm 1$ terms in the spherical harmonic expansion.
Expressing the spherical harmonics by associated Legendre polynomials, equation~(\ref{E:xi_g}) 
expands as

\begin{equation} \label{E:xi_thetaphi}
   \xi(\theta, \varphi) = -1 {2 \ell +1 \over 4 \pi} {(\ell-1)! \over(\ell+1) } 
              \sum_\ell P_\ell^1 \biggl[ \xi_\ell^1 e^{i \varphi} - \xi_\ell^{-1} e^{-i \varphi} \biggr]
,
\end{equation}
where we used the identity

\begin{equation}
   P_\ell^{-m} = (-1)^m {(\ell-m)! \over(\ell+m) }\, P_\ell^m
.
\end{equation}
In order to obtain a real function, the imaginary terms must vanish, therefore we must have 

\begin{equation} \label{E:xi_ell}
   \xi_\ell^{-1} = - \xi_\ell^1
.
\end{equation}
Using orthogonality, expressing the spherical harmonics in terms of associated Legendre polynomials
and using equations~(\ref{E:xi_ell}) and (\ref{E:xi_thetaphi}), we obtain

\begin{equation}
   \xi_\ell^1 = - {k_\ell \over b_{max} } \, \int_{\Omega_{{\hat x}^\prime}} \,
                                                     d \Omega_{{\hat x}^\prime}
                                            {g(d_c, p) \over d_c} P_\ell^1(\cos \theta) \cos^2 \varphi
,
\end{equation}
where

\begin{equation}
  k_\ell  =  \sqrt{ 2 \ell + 1 \over 4 \pi} \sqrt{ ( \ell - 1 )! \over  ( \ell + 1 )! } 
.
\end{equation}
The $\varphi$ integral can be performed 
since $g(x, p)$ and $P_\ell^1$ do not depend on $\varphi$ giving

\begin{equation}
    \xi_\ell^1 = - {\pi k_\ell \over b_{max} } \, 
                                      \int_{-1}^1\, d \mu {g(d_c, p) \over d_c}  P_\ell^1(\mu) 
,
\end{equation}
from which we find

\begin{equation} \label{E:xi_ell_1_J1}
   \xi_\ell^1 = - {\pi k_\ell \over b_{max} }   (\ell + 1/2) \,\theta_c 
                      \int_0^{p \theta_c} g(\theta, p) J_1[(\ell + {1 \over 2}) \,\theta ] \, d \theta
.
\end{equation}
Here we used the small angle approximation for the
associated Legendre polynomials:

\begin{equation}
   P_\ell^1(\mu) \approx (\ell + 1/2)\, J_1[(\ell + {1 \over 2}) \,\theta ]
,
\end{equation}
where $J_1$ is the Bessel function of the first kind and order 1 
(for a derivation see Appendix). Thus the Laplace series of the MCG effect is 

\begin{equation} \label{E:Delta_MCG_Y}
   \Delta_M = \Delta_M^{max} 
              \sum_\ell \xi_\ell^1 \Bigl( Y_\ell^1 - Y_\ell^{-1} \Bigr)
,
\end{equation}
where $\xi_\ell^1$ is given by equation~(\ref{E:xi_ell_1_J1}).

For the power spectrum of the MCG effect, using equation~(\ref{E:Delta_MCG_Y}), 
we obtain

\begin{equation}
  G_\ell^M =  { 2 \over 2 \ell + 1} \bigl( \Delta_M^{max} \bigr)^2 |\xi_\ell^1|^2
.
\end{equation}

The observed effects are calculated by convolving the theoretical fluctuation pattern
with the telescope's point spread function (PSF).
One advantage of using the spherical harmonic coefficients is that this convolution is just a 
multiplication in spherical harmonic space.
Assuming an axially symmetric PSF, $R$, its Legendre polynomial expansion may be expressed as

\begin{equation} \label{E:PSF_r}
     R({\hat x} \cdot {\hat x}^\prime) = 
                \sum_\ell  \, {2 \ell +1 \over 4 \pi } R_\ell P_\ell ({\hat x} \cdot {\hat x}^\prime) 
,
\end{equation}
where the unit vectors, $\hat x$ and ${\hat x}^\prime$, point to an arbitrary direction
(where we want to evaluate the expansion) and to the center of the PSF.
Assuming a non-axially symmetric effect, its spherical harmonic expansion can be written as

\begin{equation}
     f({\hat x}) =  \sum_{\ell , m} f_\ell^m \, Y_\ell^m({\hat x}) 
,
\end{equation}
where $\ell$ runs from zero to infinity and $m$ runs from -$\ell$ to $\ell$. 
Using the addition theorem for spherical harmonics and their orthogonality, 
the convolution of these two functions, $M = R \star f$, becomes

\begin{equation}
    M({\hat x}) = \sum_{\ell , m}  R_\ell  f_\ell^m Y_\ell^m ({\hat x})  
.       
\end{equation}

\section{Simulation of clusters of galaxies}
\label{s:Simulation}

We used Monte Carlo simulations to generate an ensemble of clusters of galaxies with masses
sampled from the PSMF (equation~\ref{E:PSMF}) with parameters those of 
our OCDM, \L CDM and SCDM models. 
We obtained the central electron number density and temperature, 
and the cluster core radius from scaling relations 
(equations~\ref{E:n_el_0}, \ref{E:T_eOM1}, \ref{E:T_eOPEN}, and \ref{E:r_cor}).

We choose to sample the PSMF using a rejection method. 
The magnitude of the peculiar velocity may be sampled using an inversion method on the 
Maxwellian (\ref{E:P_v_pec}), and yields

\begin{equation} \label{E:v_p_RN}
   v_p =  \sqrt{2} \sigma_p  \gamma^{-1} (3/2, RN)^{-1}
,
\end{equation}
where  $\gamma^{-1}(order, x)$ is the inverse of the incomplete gamma function and 
$\sigma_p$ can be determined by using equations~(\ref{E:sigm_v_pec}), 
(\ref{E:sigma_p_L0}) and (\ref{E:sigma_p_LAM}). 
$RN$ is a uniformly distributed random number in (0,1). 
We assumed an isotropic distribution in space for the directions of the peculiar 
velocity vectors, and ignored correlations between cluster peculiar velocities. 
The tangential and radial peculiar velocities are distributed as projections of 
equation~(\ref{E:v_p_RN}) accordingly.
As an illustration, in Figure~\ref{F:MCG1} we show results from one simulation using our SCDM
model projected on a grid.

The observational mass function (Figure~\ref{F:PSMF_OBS}) is specified by $M_{1.5}$, the 
mass contained within co-moving radius of 1.5 Mpc. To convert $M_{1.5}$
to the virial mass, $M_v$, which we use in the PSMF, we assume that the mass profile near 1.5 Mpc
can be approximated with $M_v(R) \propto R^q$. We obtain

\begin{equation}  
  M_v = \Biggl(   { \Delta_c(0)  \over \Delta_c \Omega_m(z) }
                           { M_{1.5} \over \Delta_c(0) (4 \pi/3) \rho_c(0) (1.5 h^{-1} {\rm Mpc})^3}
            \Biggr)^{q \over 3 - q} M_{1.5}
,
\end{equation}
or, substituting the numerical values, 

\begin{equation}  
  M_v = \Biggl(   { 178  \over \Delta_c \Omega_m(z) }
                           { M_{1.5} \over 6.98 \times 10^{14} h^{-1} M_\odot }
            \Biggr)^{q \over 3 - q} M_{1.5}
.
\end{equation}
We used q = 0.64 (\cite{Carlberget97}) to obtain curves shown in Figure~\ref{F:PSMF_Z_OBS}.

The power spectrum for an ensemble of clusters may be determined by summing the individual 
contributions of the simulated clusters (equation~[\ref{e:c_ell_int}]).
We binned clusters by their apparent core radii, $\theta_c$, then we summed the amplitudes in each bin. 
The numerical evaluation of integral (\ref{e:c_ell_int}), in this case, can be performed as

\begin{equation}
  C_\ell^{SZ} =  { 1 \over 4 \pi} \sum_i (\zeta_\ell)_i^2 
                            \sum_{cl} (\Delta_{SZ}^0)_{cl}^2
,
\end{equation}
and 

\begin{equation}
  C_\ell^M = { 2 \over 2 \ell + 1}  \sum_i (\xi_\ell^1)_i^2
                          \sum_{cl} (\Delta_M^{max})_{cl}^2
,
\end{equation}
where the index $cl$ runs over clusters whose core radii fall within the $i^{th}$ bin.

\section{Results}
\label{s:Results}

Our results for the power spectra
(more exactly the dimensionless ${\cal C}(\ell) \equiv \ell(\ell+1) C_\ell$)
from our {\it Model 1} (OCDM) are shown on Figure \ref{f:PowerSpOCDM}.
Figures~\ref{f:PowerSpLCDM} and \ref{f:PowerSpSCDM} show
our results for {\it Model 2}, (\L CDM) and {\it Model 3}, (SCDM) respectively.
As a comparison, in each figure, we plot the corresponding primordial CMBR 
power spectrum (solid line)  with $COBE$ normalization including the effects of gravitational lensing
calculated by using a new version of CMBFAST (\cite{ZaldarriagaSeljak98}; \cite{ZaldarrSperSel97}).
On large angular scales (up to about $\ell \approx 10$) the 
cosmic variance dominates (not shown).
On small angular scales the shape of the power spectra depends on the apparent angular
sizes of the clusters and the amplitudes of the effects. 
The apparent angular size depends on how the core radius and the angular diameter
distance change with redshift, while the amplitude is sensitive to the gas content, 
gas temperature and total mass as a function of redshift. 
Figures \ref{f:PowerSpOCDM}-\ref{f:PowerSpSCDM} 
demonstrate that for small angular scales ($\ell \gtrsim 3000$)  
the contribution to the power spectrum from the SSZ effect exceeds that of the primordial CMBR
in all our models. 
The contributions of the KSZ and MCG effects become important only on small scales, 
but, at those scales, they may dominate over the lensed primordial fluctuations.

Due to the early structure formation, there are more clusters at high redshift in our 
OCDM and \L CDM than in our SCDM simulations. 
Therefore the contribution to the power spectrum from clusters in a low matter density model is 
substantially larger than in a SCDM model.  
Also, most clusters are closer to us in a SCDM model, thus the contribution from clusters to the 
power spectrum peaks at higher angular scales (lower $\ell$) than in low matter density models. 
The KSZ and MCG effects have their coherence length (peak contributions) at $\ell \gtrsim 10000$.
The coherence length of the SSZ effect is about $\ell \approx 1000$ for our SCDM model and 
at about $\ell \approx 2000$ for our OCDM and \L CDM models. 
In general, the contributions to the power spectrum from the SSZ effect are
about 2 and 3 orders of magnitude greater than those from the KSZ and MCG effects.
At very small angular scales, $\ell > 10000$, the contribution to the power spectrum from the 
MCG effect might exceed that of the SSZ or KSZ effects, and even the primordial
fluctuations in the CMBR, but this depends on the details of the evolution of cluster atmospheres.

Our simulations give somewhat different results for the KSZ and MCG effects than those
of Aghanim et al. (1998).
In our simulations the amplitudes of the KSZ and MCG effects for low matter density 
models are about an order of magnitude greater than those for our SCDM model, and have a coherence 
length of about $\ell = 10000$, while rising monotonically at smaller $\ell$.
According to Aghanim et al.'s simulations, with similar cut off to ours, 
$M_{low} = 10^{13} M_\odot$, 
contributions from the KSZ effect in all models constantly grow and show no signs
of leveling off, and their amplitude has a very weak dependence on cosmological model.
Contributions from the MCG effect on the other hand show a plateau in all Aghanim et al.'s models 
for $\ell > 1000$, and for the SCDM model, the MCG effect is larger than for the other two models.

Quantitatively, our models show cluster-related effects that are weaker by 
a factor of 10 for the MCG effect in our SCDM model and a factor of 100 
for the KSZ effect for all models.
We attribute these differences mostly to the different evolution models for the intracluster gas. 
The ratio between the overall amplitudes of the KSZ and MCG effects in our calculations is
about the same as in Aghanim et al. (1998)'s results.
Our results show that the power spectra of the KSZ and MCG effects do not exceed the 
gravitationally lensed primordial power spectrum up to $\ell \approx 10000$, while the power 
spectrum of the SSZ effect becomes dominant at $\ell \gtrsim 7000$ in all our models.
Contributions to the power spectrum from the SSZ and KSZ effects 
based on Aghanim et al.'s model would exceed those from the CMBR at $\ell \gtrsim 5000$
even if one takes gravitational lensing of the primordial CMBR into account.
Our results are comparable to those obtained by Tuluie et al. (1996) and Seljak (1996).
Persi et al. (1995)'s results for contributions to the power spectrum from the SSZ (KSZ) 
effect are about the same as (an order of magnitude higher than) our results suggest.

\section{Discussion}
\label{s:Discussion}

An observed power spectrum is made up from the sum of all astrophysical effects and noise.
We rely on the different frequency and/or power spectra of the secondary effects to separate 
these foregrounds from the primordial CMBR signal (\cite{Tegmark98}).

\begin{figure} 
\centerline{
\plotone{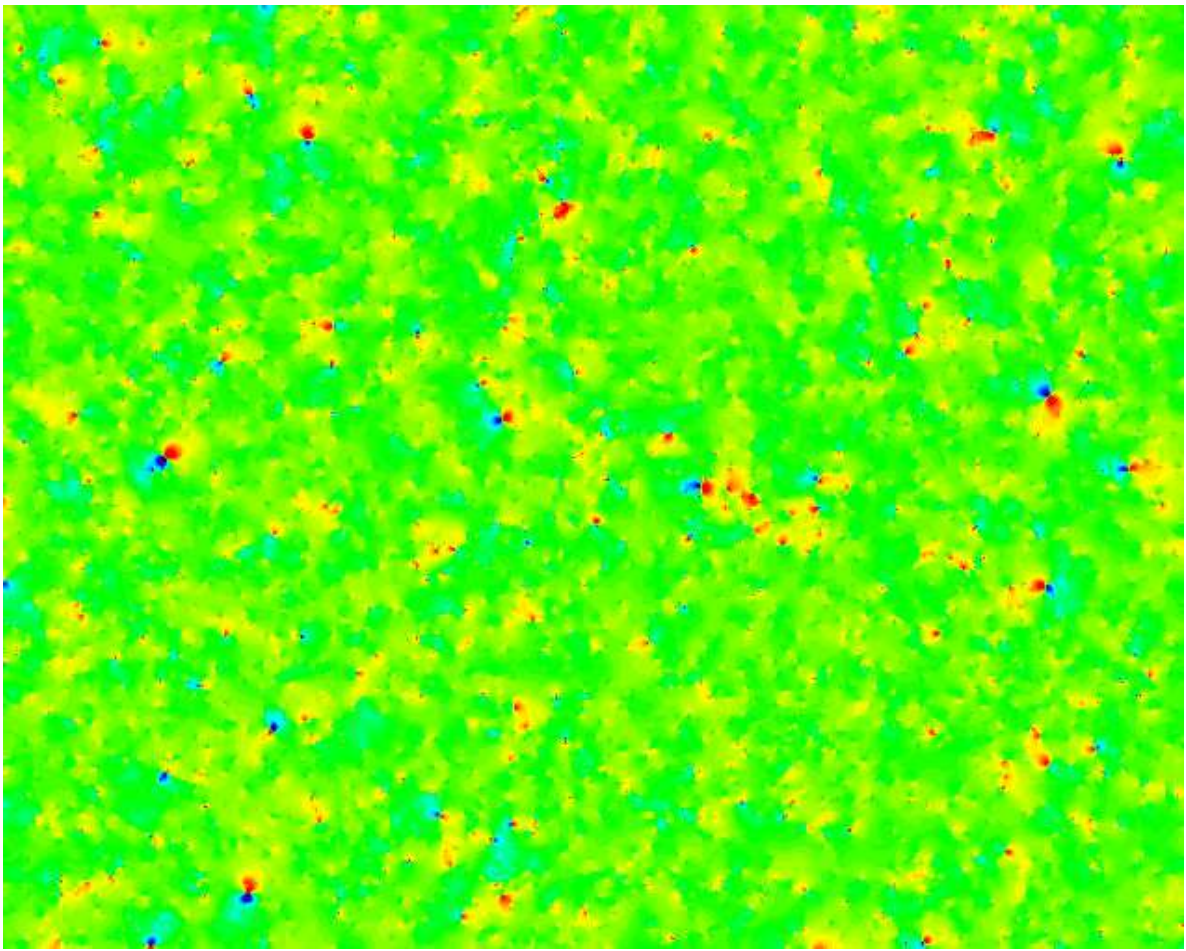}
}
\caption{
  Fluctuations in the CMBR ($\rm \Delta T$ in $\mu$K) caused by the moving 
  clusters effect in our SCDM model (see text for details).
  Primordial fluctuations and other source of secondary fluctuations are not displayed.
  The area covers $30^\circ \times 25^\circ$ of the sky with pixel size 
  $2.6^\prime \times 2.6^\prime$. The maxima and minima are about $\pm 2\mu$K.
\label{F:MCG1}
} 
\end{figure}

Of the effects discussed here, it should be easy to separate the SSZ effect
by using multi-frequency measurements of its unique spectrum. 
The separation of primordial fluctuations in the CMBR and fluctuations caused by 
the KSZ and MCG effects is more difficult since their frequency spectra are the same. 
Optimal filters have been designed to separate the
KSZ effect (\cite{HaehneltTegmark96}; \cite{Aghanimet97}): here it helps to know the SSZ effect 
for the same cluster, since that would give us a position and even an estimate for the expected
amplitude of the effect. 
Aghanim et al. (1998) discussed methods to separate the MCG effect: this is facilitated by its
unique dipole pattern with sharp peaks (Figure~\ref{F:MCG1}). 
Primordial fluctuations are usually assumed to be gaussian, where the probability of getting such 
a strongly peaked bipolar pattern is small, and we would expect the strong small angular scale 
gradient near a known cluster of galaxies to be a definite indication of the presence of the 
MCG effect. 
Also, knowing the position of clusters helps to find the effect.
However, contributions from other effects, such as early ionization and discrete radio sources
causes further confusion, and may be expected to make it difficult to determine
the power from the SZ or MCG effects.

We analyzed the contributions to the power spectrum from the SSZ, KSZ and MCG 
effects to check their impact on the determination of cosmological parameters, especially at
large $\ell$ where gravitational lensing may break the geometric degeneracy.
In Figure~\ref{f:Lens01} we show the small scale lensed primordial fluctuation power spectra 
of our three models (OCDM, SCDM, \L CDM; solid lines) with power spectra resulting 
from the sum of fluctuations due to the lensed primordial CMBR and the SSZ effect (long dashed lines), 
and from the sum of the lensed primordial CMBR, the KSZ and the MCG effects (short dashed lines). 
According to our models, if the fluctuations due to the SSZ effect are fully separated, 
the KSZ and MCG effects do not prevent the use of this part of the power spectrum to
break the geometric degeneracy and distinguish between different CDM models. 
Note that normalization at the first Doppler peak, rather than the usual {\it COBE} normalization,
would lower the contributions to the power spectrum from primordial
fluctuations in a \L CDM model relative to those from a SCDM model, and thus
secondary effects would become more important relative to the primordial CMBR fluctuations.

Our simulations also show that the power spectrum of the SSZ effect may itself be used to break the 
geometric degeneracy.
Since the separation of the SSZ effect from other secondary effects should be straightforward, 
we should be able to determine the power spectrum of the SSZ effect alone.
As can be seen from Figure~\ref{f:PowerSp3}, this power spectrum depends on 
$\Omega_0 h^2$ and $\Omega_b h^2$, providing an additional constraint on these parameters.
We note, however, that the amplitude of the SSZ effect is model dependent.
Since the contributions to the power spectrum from the SZ and MCG effects are model dependent, 
to evaluate fully their power spectra we need a better observationally-supported model for the 
intracluster gas.
Sensitive, high-resolution all-sky, X-ray observations could map the emission from intracluster gas
up to high redshift providing strong constraints on gas formation and evolution
and thus a good basis for modeling the SSZ and KSZ effects (\cite{Jahoda.et97}).
Number counts of clusters based on their SSZ effect can also be used to constrain
cosmological models (\cite{sutoet99}).

There are many possibilities of using observations to break the geometric degeneracy.
For example measurements of the CMBR polarization, the Hubble constant, or light curves of Type Ia 
supernovae have been discussed (Zaldarriaga et al. 1997; \cite{EisensteinHuTeg98}; \cite{Tegmarket98}).
Also, combination of measurements of the SSZ effect and thermal bremsstrahlung (X-ray) emission 
from clusters can be used to determine the Hubble constant for a large number of clusters, 
providing a statistical sample which might enable us to determine the Hubble constant, and
perhaps the acceleration parameter, to good accuracy (\cite{Birk98}).

Secondary fluctuations introduce non-gaussianity into the primordial spectrum
at small scales. This non-gaussianity should be taken into account when estimating
CMBR non-gaussianity at these scales. 
Winitzky (1998) estimated the effect of lensing and concluded that Planck may 
observe non-gaussianity due to lensing near the angular scale of maximum effect, $\sim 10^\prime$.
Other processes, including the SSZ, KSZ, and especially the MCG effect, 
introduce a highly non-gaussian signal as is easily seen for the MCG effect on Figure~\ref{F:MCG1}.
A similar non-gaussian pattern arises from moving cosmic strings 
(the Kaiser-Stebbins effect, compare our Figure~\ref{F:MCG1} to Figure 6a of \cite{MagueijoLewin97}).
Our results indicate that at $\ell \gtrsim 10^4$ the MCG effect might be comparable in strength 
to the primordial fluctuations. 
Evidence for non-gaussianity has been reported by Ferreira, Magueijo \& Gorski (1998) and 
Gaztanaga, Fosalba, \& Elizalde (1997) at angular scales $\ell \approx 16$ and $\ell \approx 150$.
They do not exclude the possibility that this non-gaussianity has been introduced by
foregrounds, but our results show that clusters can not introduce detectable non-gaussianity
on such scales (Figure~\ref{f:PowerSp3}).

\begin{figure} 
\centerline{
\plotone{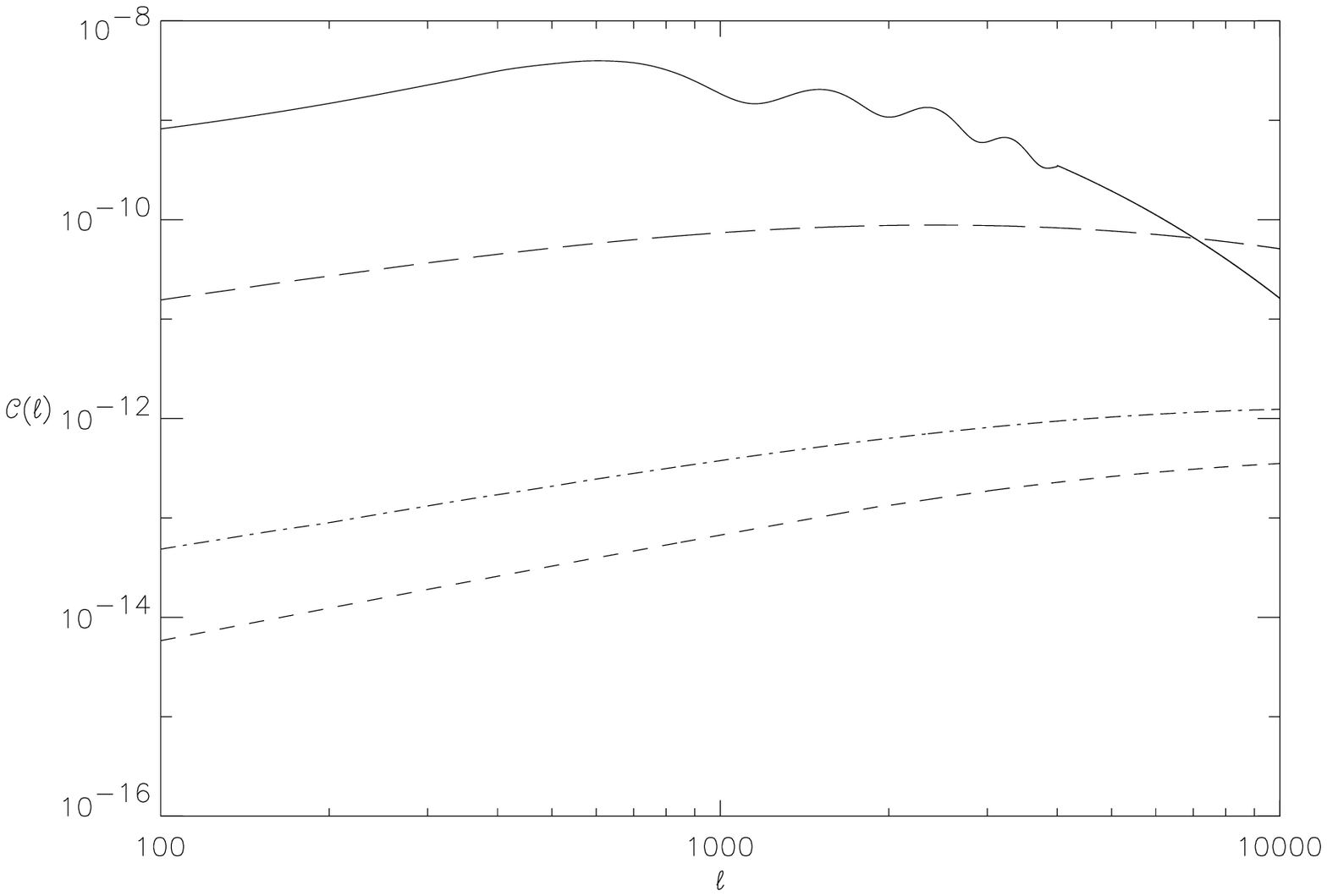}
}
\caption{ 
   The power spectrum ${\cal C}(\ell) \equiv \ell(\ell+1) C_\ell$ of the 
   lensed primordial CMBR (solid line), SSZ (long dashed), 
   KSZ (dashed dot), and MCG (short dashed) effects
   for an OCDM model with $\Omega = 0.2$, $\Lambda = 0$, $n = -1.4$.
\label{f:PowerSpOCDM}   
}
\end{figure}

\begin{figure} 
\centerline{
\plotone{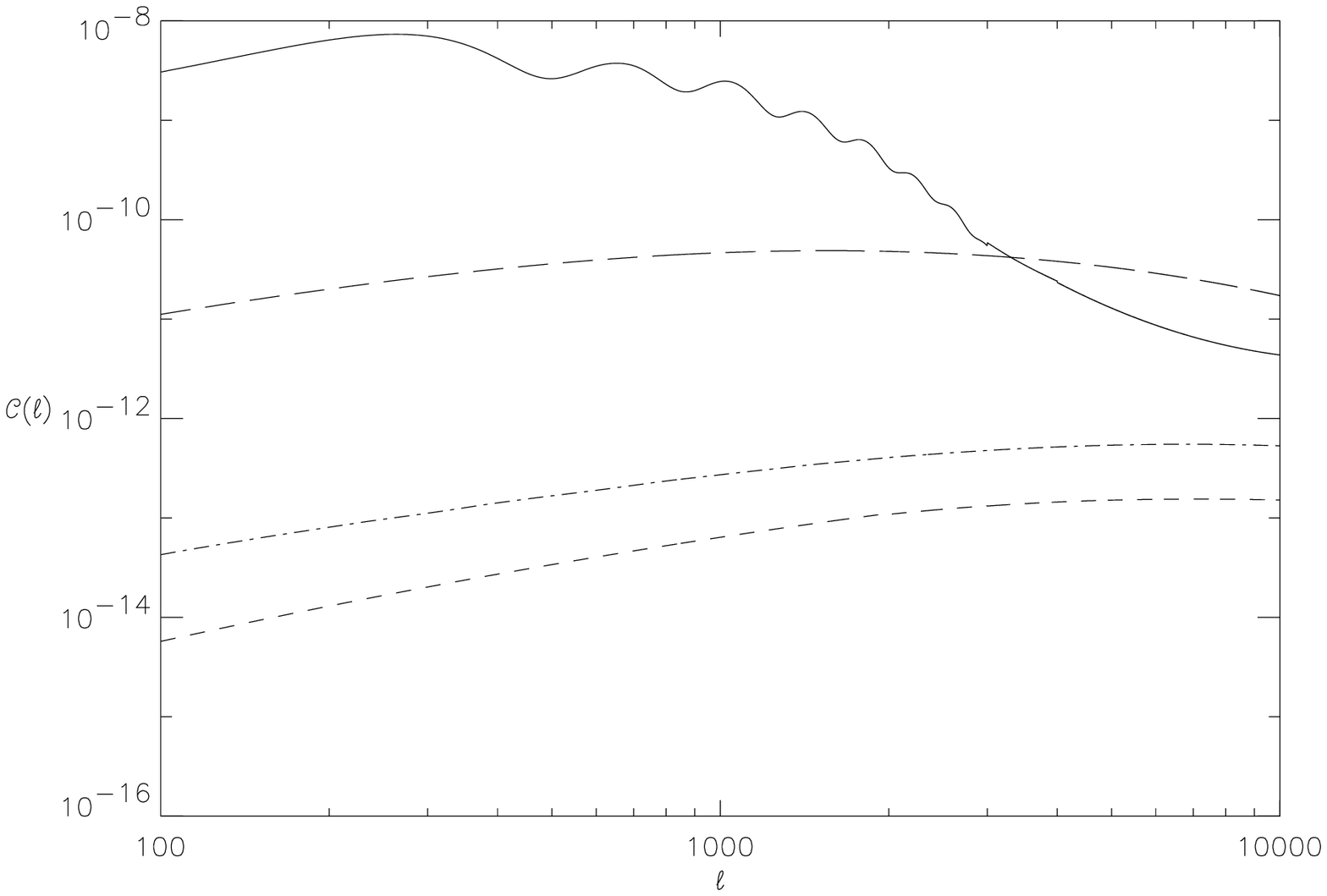}
}
\caption{
   The power spectrum ${\cal C}(\ell) \equiv \ell(\ell+1) C_\ell$ of the 
   lensed primordial CMBR (solid line), SSZ (long dashed), 
   KSZ (dashed dot), and MCG (short dashed) effects
   for a \L CDM model with $\Omega = 0.2$, $\Lambda = 0.8$, and $n = -1.4$.
\label{f:PowerSpLCDM}
}
\end{figure}

\begin{figure} 
\centerline{
\plotone{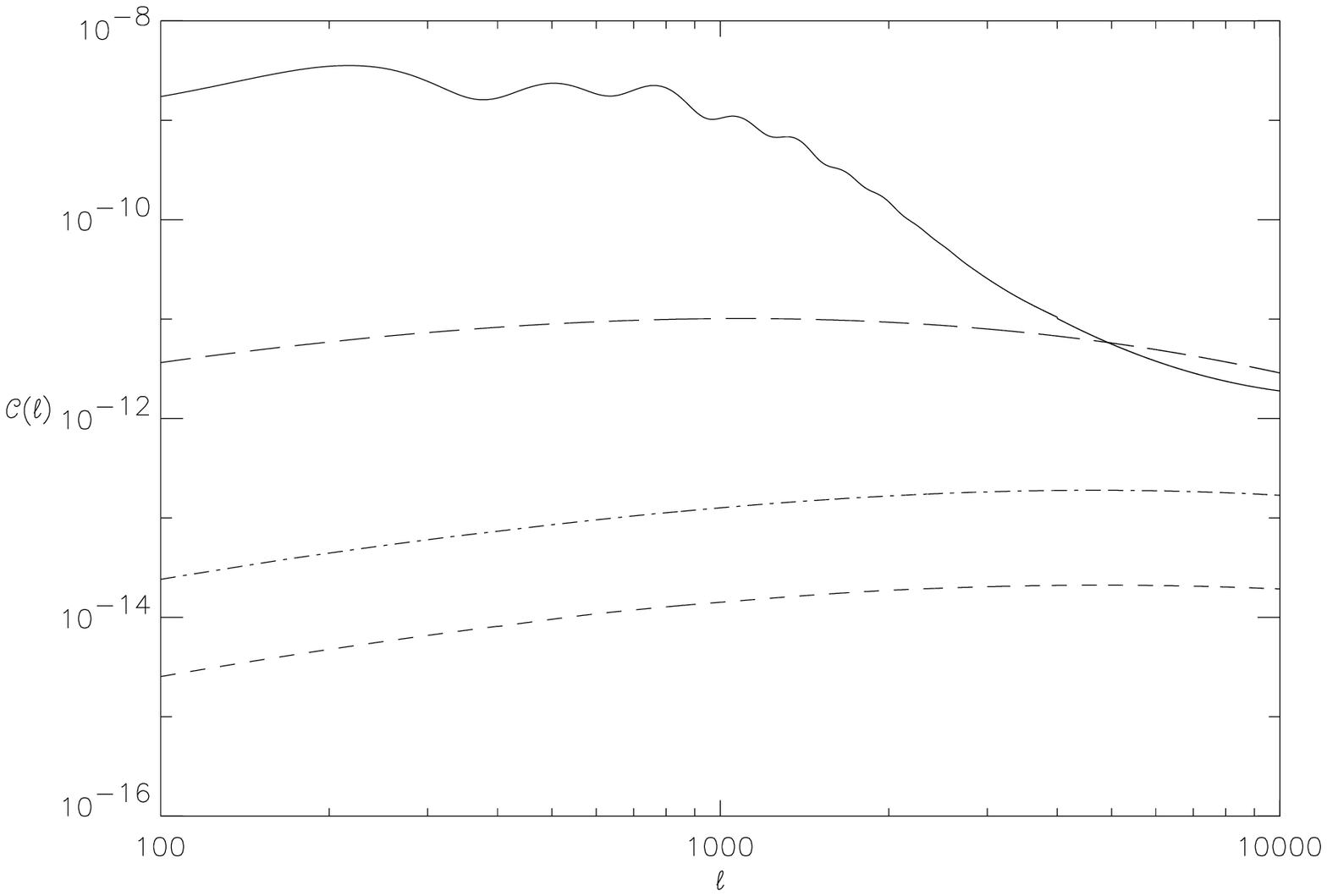}
}
\caption{
   The power spectrum ${\cal C}(\ell) \equiv \ell(\ell+1) C_\ell$ of the 
   lensed primordial CMBR (solid line), SSZ (long dashed), 
   KSZ (dashed dot), and MCG (short dashed) effects
   for an OCDM model with$\Omega = 1$, $\Lambda = 0$, and $n = -1.4$.
\label{f:PowerSpSCDM}
}
\end{figure}

\begin{figure} 
\centerline{
\plotone{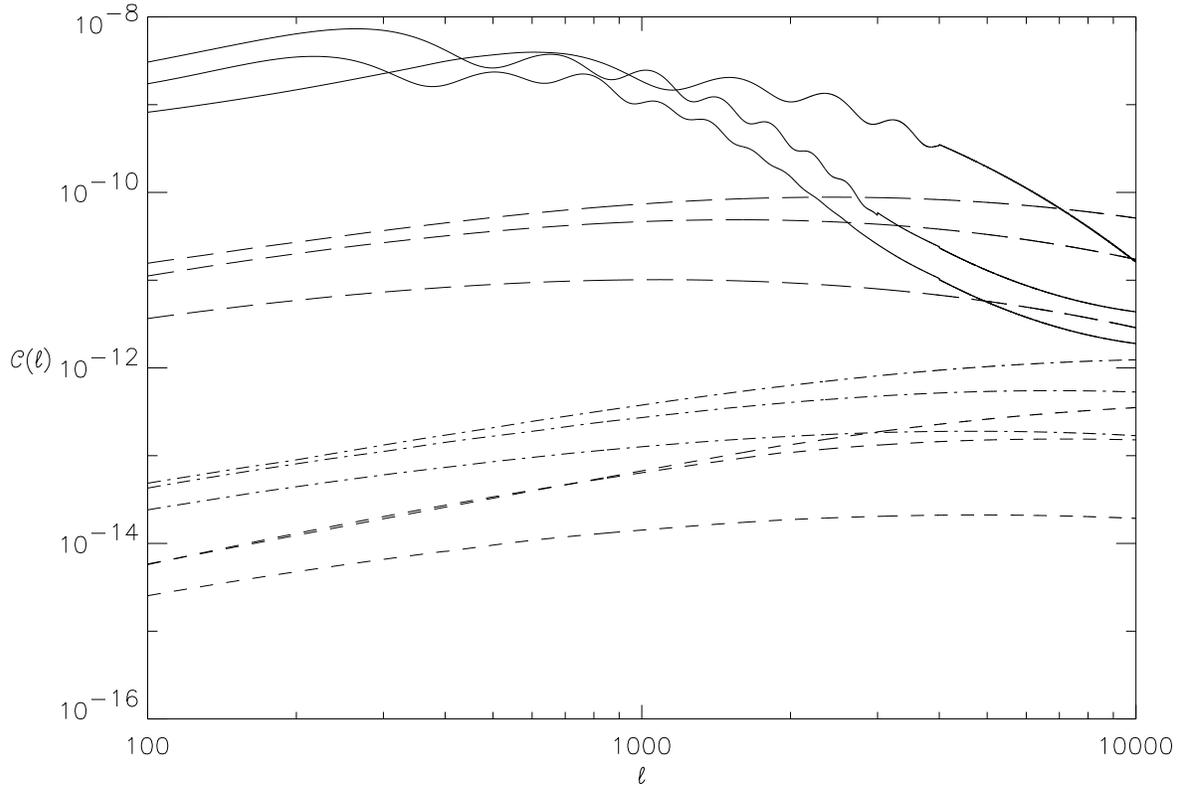}
}
\caption{
  Power spectra ${\cal C}(\ell) \equiv \ell(\ell+1) C_\ell$: 
  lensed primordial CMBR (solid line), SSZ (long dashed), 
  KSZ (dashed dot), and MCG (short dashed) effects
  for our three models: OCDM, \L CDM, and SCDM.
\label{f:PowerSp3}
}
\end{figure}

\begin{figure} 
\centerline{
\plotone{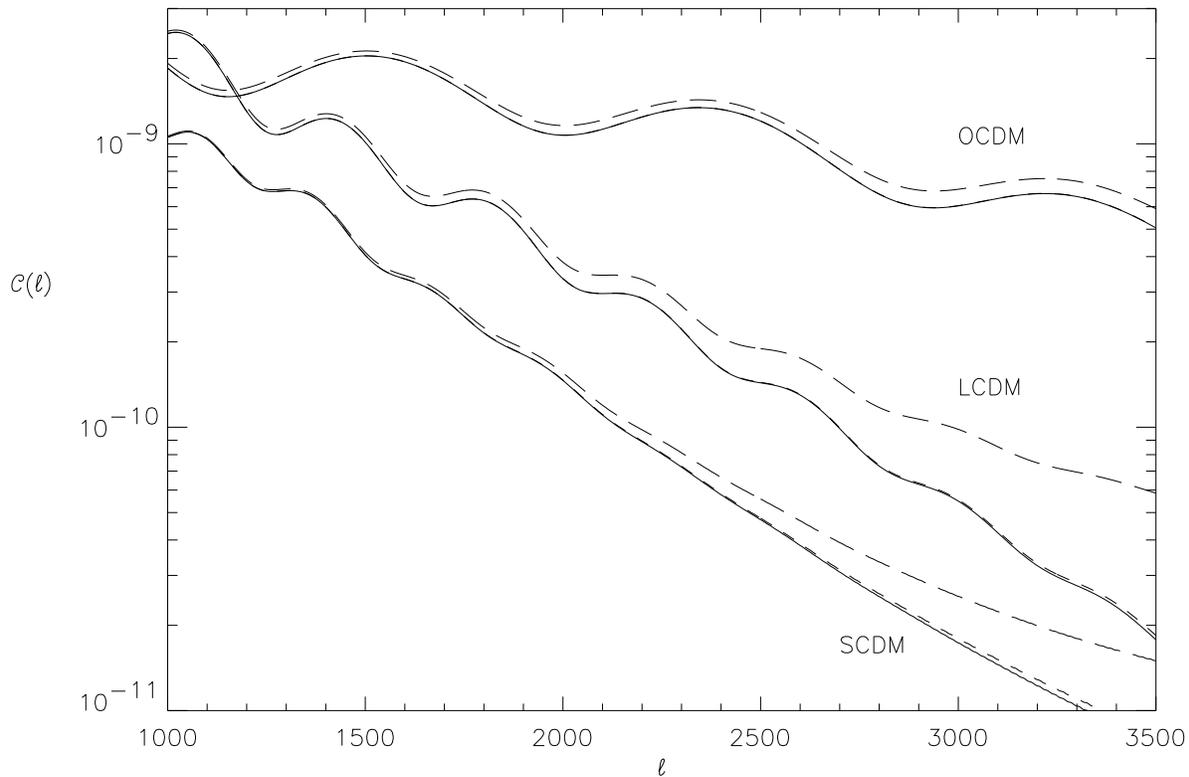}
}
\caption{
  Small scale power spectrum ${\cal C}(\ell) \equiv \ell(\ell+1) C_\ell$
  of the lensed primordial CMBR, 
  the static and kinematic SZ effects, and the MCG effect for our three
  models: OCDM, \L CDM, SCDM. The long and short 
  dashed lines represent the sum of the lensed primordial CMBR and the SSZ effect, and the sum
  of the lensed primordial CMBR, the KSZ and RSC effects respectively. 
  The contribution from the static SZ effect is important in all models. 
  Contributions from the kinematic SZ and MCG effects are negligible at these $\ell$ values.
  We may conclude that if fluctuations due to the
  static SZ effect are removed the kinematic and MCG effects do not
  prevent using gravitational lensing to break the geometric degeneracy. 
  Note that different normalizations shift the sets of curves.
  For example, normalization to the first Doppler peak makes
  the power spectra of our SCDM and \L CDM models shift closer.
\label{f:Lens01}
}
\end{figure}

We convolved our theoretical results (Figures~\ref{f:PowerSpOCDM} - \ref{f:PowerSpSCDM}) with the 
expected point spread functions (PSFs) of instruments on the MAP and Planck missions to estimate the 
level of the secondary fluctuations caused by clusters of galaxies on the observable power spectrum.
The observed $C_{\ell}$ values become

\begin{equation}
    C_{\ell}^{obs} = C_{\ell} W_{\ell}
,
\end{equation}
where the $W_{\ell}$ values are the Legendre coefficients of the PSF.
For an assumed gaussian response function and in the small angle approximation, the 
$W_{\ell}$ coefficients are

\begin{equation}
    W_{\ell} = e^{- \sigma^2 (\ell + 1/2)^2}
,
\end{equation}
where $\sigma = h / 2 \sqrt{ \ln 4}$, and $h$ is the FWHM
of the beam (for a detailed description of window functions and $W_{\ell}$ coefficients, 
see White \& Srednicki 1995).
The observed rms fluctuations then become 

\begin{equation} \label{E:DT_rms}
   \langle \Delta T/ T_0 \rangle_{rms}^2 = 
                        \sum_\ell { 2 \ell + 1 \over 4 \pi}  C_\ell W_{\ell} 
.
\end{equation}

In general, contributions from unresolved cluster static effects add to provide a cumulative 
contribution to the CMBR power spectrum. Contributions from the KSZ and MCG
effects from unresolved sources tend to cancel. In the case of the MCG effect this is because
each unresolved source contribution would be zero owing to the dipole spatial
pattern of the effect. For small-scale KSZ effects there are several
sources in the field of view of the telescope, and different sources have positive or
negative contributions depending on the sign of their line of sight peculiar velocity, and
therefore they tend to cancel each other.
The larger the beam size, the more effective is the cancellation of the MCG and KSZ effects.
Note that the spatial extension of the MCG effect is much
larger than that of the KSZ effect, so many clusters may be unresolved in their KSZ
and resolved in their MCG effect.
The MCG effect might be relatively more important at high redshifts, since it does not require
a well-developed cluster atmosphere.

\begin{figure} 
\centerline{
\plotone{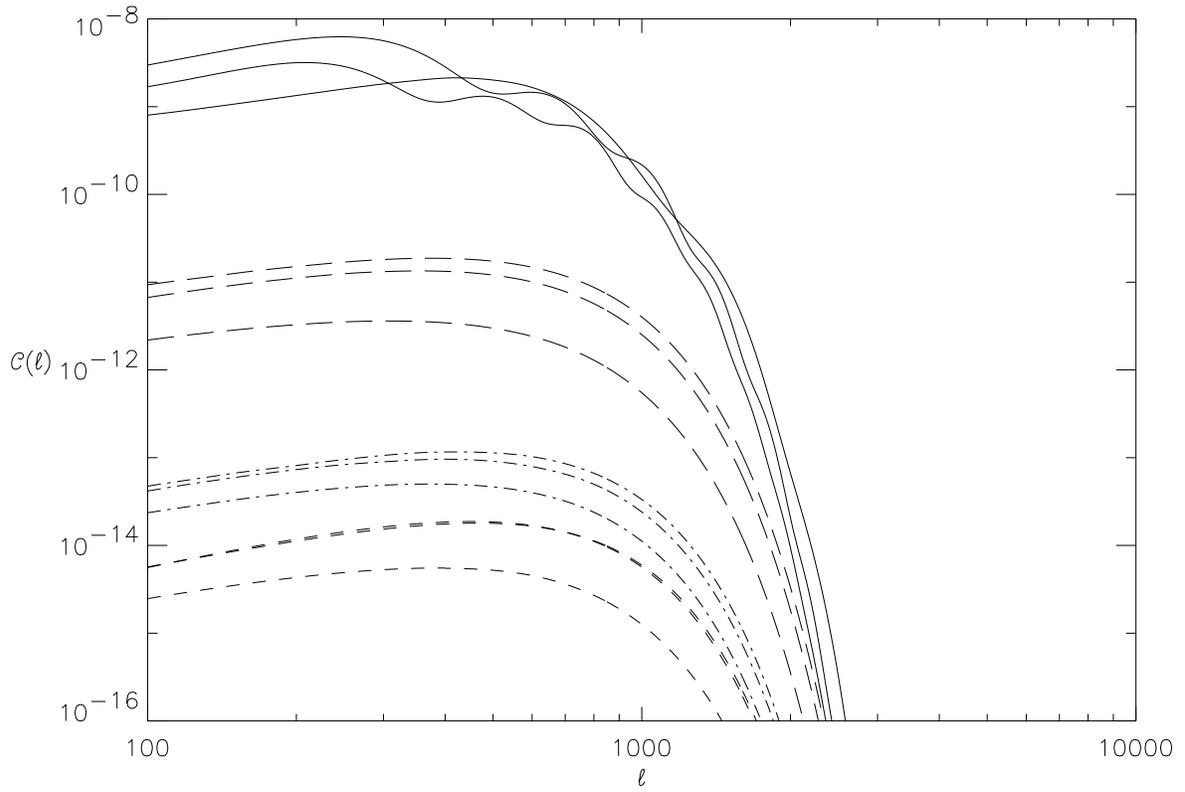}
}
\caption{
  Power spectra ${\cal C}(\ell) \equiv \ell(\ell+1) C_\ell$
  of fluctuations in the CMBR as MAP would
  observe them at $\nu = 94$ GHz with FWHM = 12.6$^\prime$.
  Models and line codes are the same as in Figure~\ref{f:PowerSp3}.
  The contribution from the primordial fluctuations dominates on all scales.
\label{f:MAP94}
}
\end{figure}

\begin{figure} 
\centerline{
\plotone{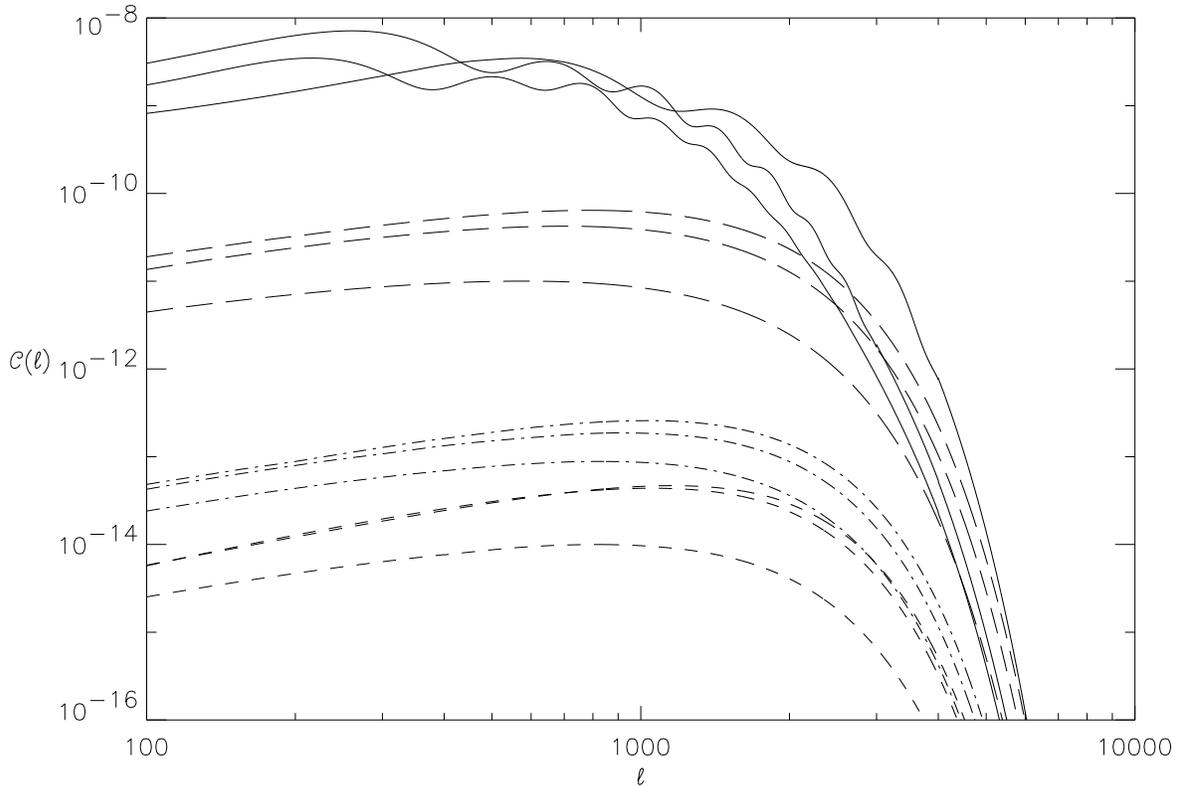}
}
\caption{
  Power spectra ${\cal C}(\ell) \equiv \ell(\ell+1) C_\ell$
  of fluctuations in the CMBR as Planck would
  observe them at $\nu = 353$ GHz with FWHM = 5 arc minute.
  Models and line codes are the same as in Figure~\ref{f:PowerSp3}.
  The contribution to the power spectrum from the static SZ effect 
  at this frequency is 1.2 times that in the Rayleigh-Jeans region, 
  and the static SZ effect dominates over the processed primordial fluctuations
  above about $\ell \approx 3000$ (except for the SCDM model).
\label{f:PLANCK353}
}
\end{figure}

In Figures~\ref{f:MAP94} and \ref{f:PLANCK353} we show the
contributions to the power spectrum from primordial fluctuations, and the SSZ, KSZ and MCG effects, 
convolved with the PSF of the planned $\nu = 94$ GHz receiver on MAP, and the planned 
$\nu = 353$ GHz bolometer on Planck.
The amplitude of the fluctuations from the the SSZ effect is negative at $\nu = 94$ GHz 
and positive at $\nu = 353$ GHz, but only the the absolute value of the effect contributes to 
the power spectrum.
The different maximum $\ell$ values for the MAP and Planck systems ($\ell_{max} \sim$ 1000 and 2000,
respectively) can clearly be seen on Figures~\ref{f:MAP94} and \ref{f:PLANCK353}.
Because of these cutoffs, the observable power spectrum is dominated by 
primordial fluctuations at all $\ell$ for these missions.
According to our results, the SSZ effect may cause a 1\% 
enhancement in the amplitude of the Doppler peaks, which is at the limit of the sensitivity
of the MAP and Planck missions.
From the analysis of the power spectrum, this would lead to an overestimation
of the parameter $\Omega_0 h^2$ by about 1\%,.
% !!!!!!    of the parameter $\Omega_0 h^2$ by about 1\%.
The shift in the position of peaks as a function of $\ell$ caused by the SSZ effect 
is less important since 
the spectrum of the SSZ effect has only a weak dependence on $\ell$.

In Table~\ref{T:RMS_ALL} we show the $(\Delta T/T)_{rms}$ values of the contributions to the CMBR
from the SSZ, KSZ, and MCG effects convolved with the the 94 GHz MAP and 353 GHz Planck
receivers for our three models (OCDM; \L CDM; SCDM).
As a comparison, we display the corresponding rms values of the primordial fluctuations.
The rms values of all these secondary effects are an order of magnitude
smaller than rms values from primordial fluctuations. The most important contribution
is that of the SSZ effect at these frequencies. The KSZ and MCG effects give similar contributions
with the KSZ effect being about a factor of two stronger. 
Aghanim et al. (1998)'s results for the rms values of the MCG effect is a factor of 10 (SCDM)
or a factor of 3 (OCDM and \L CDM) larger than our results.
Note, however, that rms values give only a crude estimate of the magnitude of the effects. 
At large angular scales the primordial fluctuations are about 100 (for the SSZ effect)
or $10^5 - 10^7$ (KSZ, MCG effects) times stronger than the secondary fluctuations. 

An ideal observation to measure the contribution to the power spectrum from the
SSZ effect would use high angular resolution ($\ell \gtrsim 7000$) and high frequency
($\nu \gtrsim 250$ GHz). 
SuZIE probes the power spectrum at angular scale $\ell \approx 7500$ at 140 GHz.
The 2$\sigma$ upper limit on the power at this scale from SuZIE is 
$\ell (\ell+1) C_\ell \leq  1.4 \times 10^{-9}$ (\cite{Gangaet97}). 
Unfortunately our models suggest that at this frequency the primordial contribution 
to the power spectrum is about 10 times stronger than that from the SSZ effect.
A promising experiment is SCUBA, which probes the anisotropies at 
angular scale $\ell \approx 10000$ and frequency 348.4 GHz.
Their preliminary 2$\sigma$ upper limit on the power spectrum at this scale is 
$\ell (\ell+1) C_\ell \leq  4.7 \times 10^{-8}$. 
Much further work is planned, and should lower this limit by a factor of 3-10 (\cite{BorysCS98}).

\acknowledgments

SMM is grateful to Bristol University for a full scholarship, where most of this work was done. 
This work was finished while SMM held a National Research Council Research Associateship at 
NASA Goddard Space Flight Center. 
We thank N. Aghanim for comments on an earlier version of the manuscript, and 
our referee, Dr Bartlett, for his detailed comments and for helping us to clarify some aspects of 
our approximations. We thank U. Seljak and M. Zaldarriaga for making the CMBFAST code available.

% % % % % % % % % % % % % % % % % % % % % % % % % % % % % % % % % % % % % % % % % % % % % % % % %
% 
%                                     A P P E N D I X       
% 
% % % % % % % % % % % % % % % % % % % % % % % % % % % % % % % % % % % % % % % % % % % % % % % % %

\clearpage

\appendix

\centerline{ \bf{Appendix: Small Angle Approximation for $P_\ell^1$}}

We derive a small angle approximation for the associated Legendre polynomial
$P_\ell^1$. We express the associated Legendre polynomial
$P_\ell^1$ in terms of the Legendre polynomial $P_\ell$
(see for example \cite{ArfkenWeber95}) as 

\begin{equation} 
  P_\ell^1(\mu) = \sqrt{1 - \mu^2} { d \over d \mu} P_\ell (\mu)
.
\end{equation}
Introducing a new variable, $x = \ell(\ell+1) \theta$, using the chain rule and a 
small angle approximation, we get

\begin{equation} 
  P_\ell^1(\mu) = \sqrt{1 - \mu^2} { d  P_\ell\over d x} { d x \over d \theta}  
                          { d  \theta\over d \mu} \approx - \ell(\ell+1) { d  P_\ell\over d x} 
.
\end{equation}
Using a small angle approximation for $P_\ell$ (Peebles 1980) we obtain

\begin{equation} 
   - \ell(\ell+1) { d  P_\ell\over d x} \approx   - \ell(\ell+1)  { d  \over d x} J_0 (x)
.
\end{equation}
Using the relation  

\begin{equation} 
          J_1(x) =  - { d  \over d x} J_0 (x) 
\end{equation}
for Bessel functions, we finally obtain

\begin{equation} \label{E:P_ell_1_J1}
   P_\ell^1(\mu) \approx   - \ell(\ell+1)  { d  \over d x} J_0 (x) = \ell(\ell+1) J_1(x)
.
\end{equation}

% % % % % % % % % % % % % % % % % % % % % % % % % % % % % % % % % % % % % % % % % % % % % % % % %
% 
%                                     B I B L I O G R A P H Y         
% 
% % % % % % % % % % % % % % % % % % % % % % % % % % % % % % % % % % % % % % % % % % % % % % % % %

% % % % % % % % % % % % % % % % % % % % % % % % % % % % % % % % % % % % % % % % % % % % % % % % %
% 
%                 F I G U R E S            F I G U R E S                 F I G U R E S           
% 
% % % % % % % % % % % % % % % % % % % % % % % % % % % % % % % % % % % % % % % % % % % % % % % % %

\clearpage

%  FIGURE 1

%  FIGURE 2

%  FIGURE 3

%  FIGURE 4

%  FIGURE 5

%  FIGURE 6

%  FIGURE 7

%  FIGURE 8

%  FIGURE 9

%  FIGURE 10

% % % % % % % % % % % % % % % % % % % % % % % % % % % % % % % % % % % % % % % % % % % % % % % % % % % 
% 
%                                              T A B L E S 
% 
% % % % % % % % % % % % % % % % % % % % % % % % % % % % % % % % % % % % % % % % % % % % % % % % % % % 

\clearpage

\begin{table}
\begin{center}
\begin{tabular}{cccccccccc} 
\hline
\hline
\noalign{\vskip 1pt} \\
  Model  &   & MAP     & 94 GHz  &       &    &          & Planck & 353 GHz &       \\
\noalign{\vskip 0pt} \\
\hline
\noalign{\vskip 0pt} \\
     &  CMBR   &   SSZ  &  PKSZ  &  MCG &  &  CMBR  &     SSZ     &  PKSZ  &  MCG  \\
     &  $\mu$K  &  $\mu$K  &  $\mu$K  &  $\mu$K  &   &  $\mu$K  &  $\mu$K  &  $\mu$K  &  $\mu$K \\
\noalign{\vskip 0pt} \\
\hline
\hline \\[-2pt]
1. OCDM  &   86  &  7.0 &  0.54 &  0.20 &  &  100 &  14  &  0.81 &  0.33    \\[-2pt]
2. LCDM  &  119  &  5.9 &  0.49 &  0.20 &  &  130 &  11  &  0.71 &  0.32    \\[-2pt]
3. SCDM  &   93  &  3.2 &  0.36 &  0.12 &  &  101 &  5.7 &  0.50 &  0.17    \\[-2pt]
\end{tabular} 
\end{center} 
\caption{\label{T:RMS_ALL}
   Root mean square ($\Delta T_{rms}$) of the SSZ, KSZ, and MCG effects convolved 
   with response functions of the 94 GHz MAP (FWHM = $12.6^\prime$) and 
   353 GHz Planck (FWHM = $5^\prime$) instruments from our three models 
   (OCDM; LCDM; SCDM, see section~\ref{S:INTRO}). 
   As a comparison, we display the rms values of the primordial fluctuations (CMBR).
        }
\end{table}

% % % % % % % % % % % % % % % % % % % % % % % % % % % % % % % % % % % % % % % % % % % % % % % % % 


\clearpage

\begin{thebibliography}{plain}

\bibitem[Aghanim et al. 1997]{Aghanimet97}
   Aghanim, N., De Luca, A., Bouchet, F. R., Gispert, R., \& Puget, J. L., 
   1997, A\&A, 325, 9

\bibitem[Aghanim et al. 1998]{Aghanimet98}
   Aghanim, N., Prunet, S., Forni, O., \& Bouchet, F. R., 1998, A\&A, 334, 409

\bibitem[Arfken \& Weber 1995]{ArfkenWeber95} 
            Arfken, G. B., \& Weber, H. J., 1995, ``Mathematical Methods for Physicists'',
            New York \& London: Academic Press

\bibitem[Atrio-Barandela \& Mucket 1998]{AtBaMuck99}
   Atrio-Barandela, F., \& Mucket, J. P., 1999, \apj, 515, 465

\bibitem[Bahcall \& Cen 1993]{BahcallCan93}
   Bahcall, N. A., \& Cen, R., 1993, \apj, 407, L49

\bibitem[Bahcall \& Fan 1998]{BahcallFAn98}
   Bahcall, N. A., \& Fan, X., 1998, \apj, 504, 1

\bibitem[Bahcall \& Oh 1996]{BahcallOh96}
   Bahcall, N. A., \& Oh, S. P., 1996, \apjl, 462, L49

\bibitem[Bahcall et al. 1997]{Bahcallet97}
   Bahcall, N. A., Fan, X., \& Cen, R. ,1997, \apj, 485, L53

\bibitem[Bardeen et al. 1986]{Bardeen86}
       Bardeen, J. M., Bond, J. R., Kaiser, N., \& Szalay, A. S., 
       1986, \apj, 304, 15.

\bibitem[Bartlett \& Silk 1994]{BartSilk94}
   Bartlett, J. G., \& Silk, J., 1994, \apj, 423, 12

\bibitem[Bersanelli et al. 1996]{Bersanelliet96} 
   Bersanelli et al., 1996, ``Report on Phase A Study for COBRAS/SAMBA, ESA

\bibitem[Birkinshaw 1998]{Birk98}
   Birkinshaw, M., 1998, Physics Reports, 310, 97

\bibitem[Birkinshaw \& Gull 1983]{BirkGull83} 
   Birkinshaw, M., \& Gull, S. F., 1983, Nature, 302, 315

\bibitem[Blanchard \& Schneider 1987]{BlanchardSchneider87}
   Blanchard, A., \& Schneider, J., 1987, A\&A, 184, 1

\bibitem[Bond, Efstathiou \& Tegmark 1997]{BondEfTeg97}
   Bond, J. R., Efstathiou, G., \& Tegmark, M., 1997, \mnras, 291. L33

\bibitem[Borys, Chapman \& Scott 1998]{BorysCS98}
     Borys, C., Chapman, S. C., \& Scott, D., 1998, preprint, astro-ph/9808031

\bibitem[Bryan \& Norman 1998]{BryanNorman98}
   Bryan, G. L., \& Norman, M. L., 1998, \apj, 495, 80 

\bibitem[Carlberg, Yee \& Ellingson 1997]{Carlberget97}
       Carlberg, R. G., Yee, H. K. C., \& Ellingson, E., 1997, \apj, 478, 46

\bibitem[Caroll et al. 1992]{Carollet92}
      Caroll, S. M., Press, W. H., Terner, E. L., 1992, ARAA, 30, 499

\bibitem[Cavaliere \& Fusco-Femiano 1976]{CavaliereF76} 
   Cavaliere, A., \& Fusco-Femiano, R., 1976, \aap, 49, 137

\bibitem[Colafrancesco \& Blasi 1998]{ColaBlasi98}
   Colafrancesco, S., \& Blasi, P., 1998, preprint, astro-ph/9804262

\bibitem[Colafrancesco \& Vittorio 1994]{ColaVittorio94}
   Colafrancesco, S., \& Vittorio, N., 1994, \apj, 422, 443

\bibitem[Colafrancesco et al. 1994]{ColaMRV94}
   Colafrancesco, S, Mazzotta, P, Rephaeli, Y, Vittorio 1994, \apj, 433, 454

\bibitem[Colafrancesco et al. 1997]{ColaMRV97}
   Colafrancesco, S, Mazzotta, P, Rephaeli, Y, Vittorio 1997, \apj, 479, 1

\bibitem[Colberg et al. 1998]{Colberg98}
    Colberg, J. M., White, S. D. M., MacFarland, T. J., Jenkins, A.,
    Pearce, F. R., Frenk, C. S., Thomas, P. A., Couchman, H. M. P., 
    1998, preprint, astro-ph/9805078

\bibitem[Cole \& Kaiser 1988]{ColeKaiser88}
     Cole, S., \& Kaiser, N., 1988, \mnras, 233, 637

\bibitem[De Luca et al. 1995]{DeLucaet95}
   De Luca, A., Desert, F. X., \& Puget, J. L., 1995, A\&A, 300, 335

\bibitem[Efstathiou \& Bond 1998]{EfstathBond98}
     Efstathiou, G., \& Bond, J. R., 1998, preprint, astro-ph/9807103

\bibitem[Eisenstein, Hu \& Tegmark 1998]{EisensteinHuTeg98} 
     Eisenstein, D. J., Hu, W., \& Tegmark, M., 1998, preprint, astro-ph/9807130

\bibitem[Eke, Cole \& Frenk 1996]{Ekeet96}
         Eke, V. R, Cole, S, Frenk 1996, \mnras, 282, 263

\bibitem[Eke et al. 1998]{Ekeet98}
         Eke, V. R, Cole, S, Frenk, S, \& Henry, J. P., 1998, \mnras, 298, 1145

\bibitem[Evrard, Metzler \& Navarro 1996]{EvrardMetNav96}
         Evrard, A. E. Metzler, C. A., \& Navarro, J. F., 1996, \apj, 469, 494

\bibitem[Ferreira, Magueijo \& Gorski 1998]{Ferreiraet98}
   Ferreira, P. D., Magueijo, J., \& Gorski, K. M., 1998, preprint, astro-ph/9803256

\bibitem[Fixsen et al. 1996]{Fixsenet96}
   Fixsen, D. J., Cheng, E. S., Gales, J. M., Mather, J. C., Shafer,
   R. A., \& Wright, E. L., 1996, \apj, 473, 576

\bibitem[Frenk et al. 1999]{Frenket99}
   Frenk et al., 1999, \apj, 525, 554

\bibitem[Ganga et al. 1997]{Gangaet97} 
   Ganga et al., 1997, preprint, astro-ph/9702186

\bibitem[Gaztanaga, Fosalba \& Elizalde 1997]{Gaztanagaet97}  
   Gaztanaga, E., Fosalba, P., \& Elizalde, E., 1997, preprint, astro-ph/9705116

\bibitem[Girardi et al. 1996]{Girardiet96}
   Girardi et al., 1996, \apj, 457, 61

\bibitem[Gradshteyn \& Ryzhik 1980]{GradshtRyzh80}
   Gradshteyn, I. S., \& Ryzhik, I. M., 1980, ``Table of Integrals Series, \& Products'',
   New York \& London: Academic Press

\bibitem[Gramann 1998]{Gramann98} 
   Gramann, M., 1998, \apj, 493, 28

\bibitem[Gramann et al. 1995]{Gramannet95} 
   Gramann, M., Bahcall, N. A., Cen, A., \& Gott, J. R., 1995, \apj, 441, 449

\bibitem[Gurvits \& Mitrofanov 1986]{GurvitsMitrof86} 
   Gurvits, L. I., \& Mitrofanov, I. G., 1983, 324, 349

\bibitem[Haehnelt \& Tegmark 1996]{HaehneltTegmark96}
   Haehnelt, M. G., \& Tegmark, M., 1996, \mnras, 279, 545

\bibitem[Hobson et al. 1998]{Hobsonet98}
   Hobson, M. P., Jones, A. W., Lasenby, A. N., Bouchet, F. R., 
   1998, preprint, astro-ph/9806387

\bibitem[Horner et al. 1999]{Horneret99}
   Horner et al 1999, preprint, astro-ph/9902151

\bibitem[Jaffe \& Kamionkowski 1998]{JaffKami98}
   Jaffe, A. H., \& Kamionkowski, M., 1998, preprint, astro-ph/9801022v3

\bibitem[Jahoda 1997]{Jahoda.et97} 
   Jahoda, K., 1997, preprint, astro-ph/9711287

\bibitem[Jones \& Forman 1984]{JonesForman84}
   Jones, C., \& Forman, W., 1984, \apj, 276, 38

\bibitem[Kaiser 1986]{Kaiser86} 
   Kaiser, N., 1986, \mnras, 222, 323

\bibitem[King 1962]{King62}
   King, I. R., 1962, \aj, 67, 471

\bibitem[Kitayama \& Suto 1996]{KitaSuto96}
   Kitayama, T., \& Suto, Y., 1996, \apj, 469, 480

\bibitem[Lacey \& Cole 1993]{LaceyCole93}
   Lacey,  C., \& Cole, S., 1993, MNRAS, 262, 627

\bibitem[Lahav et al. 1991]{lahavet91}
   Lahav, O, Rees, M. J., Lilje, P. B., \& Primack, J. R., 1991, \mnras, 251, 128

\bibitem[Lineweaver \& Barbosa 1998]{LinewBar98}
   Lineweaver, C. H., \& Barbosa, D., 1998, \apj, 496, 624 

\bibitem[Lubin \&  Bahcall 1993]{LubinBahcall93}
   Lubin, L. M., \& Bahcall, N. A., 1993, \apjl, 415, L17

\bibitem[Magueijo \& Lewin 1997]{MagueijoLewin97}
   Magueijo, J., \& Lewin, A., 1997, preprint, astro-ph/9702131

\bibitem[Makino \& Suto 1993]{MakiSuto93}
   Makino, N., \& Suto, Y 1993, \apj, 405, 1

\bibitem[Markevitch et al. 1992]{Markevitchet92} 
   Markevitch, M., Blumenthal, G. R., Forman, W., Jones, C., \& Sunyaev, R. A., 
   1992, \apj, 395, 326

\bibitem[Markevitch et al. 1994]{MarkBFJS94} 
   Markevitch, M., Blumenthal, G. R., Forman, W., Jones, C., \& Sunyaev, R. A., 
   1994, \apj, 426, 1

\bibitem[Metcalf \& Silk 1997]{MetcalfSilk97}
   Metcalf, R. B., \& Silk, J., 1997, \apj, 489, 1

\bibitem[Metcalf \& Silk 1998]{MetcalfSilk98}
   Metcalf, R. B., \& Silk, J., 1998, \apjl, 492, L1

\bibitem[Molnar 1998]{Molnar98}
   Molnar, S.M., 1998, PhD Thesis, University of Bristol

\bibitem[Molnar \& Birkinshaw 1999]{MolnarBirkinshaw99} 
   Molnar, S. M., \& Birkinshaw, M., 1999, \apj, 523, 78

\bibitem[Navarro, Frenk \& White 1997]{Navarroet97}
   Navarro, J. F., Frenk, C. S., \& White, S. D. M., 1997, \apj, 490, 493

\bibitem[Nozawa et al. 1998]{NozaIK98}
   Nozawa, S., Itoh, N., \& Kohyama, Y., 1998, \apj, 508, 17

\bibitem[Ostriker \& Vishniac 1986]{OstrVish86}
   Ostriker, J. P., \& Vishniac, E. T., 1986, \apjl, 306, L51

\bibitem[Oukbir \& Blanchard 1997]{OukbirBlanchard97}
   Oukbir, J., \& Blanchard 1997, A\&A, 317, 1

\bibitem[Persi et al. 1995]{Persiet95}
   Persi, F. M., Spergel, D. N., Cen, R., \& Ostriker, J. P., 1995, \apj, 442, 1

\bibitem[Peebles 1980]{Peebles80}
   Peebles, P. J., 1980, ``The Large Scale Structure of the Universe'', 
   Princeton: Princeton University Press

\bibitem[Peebles 1984]{Peebles84}
   Peebles, P. J., 1984, \apj, 284, 439

\bibitem[Peebles 1993]{Peebles93}
   Peebles, P. J., 1993, ``Principles of Physical Cosmology'', 
   Princeton: Princeton University Press

\bibitem[Press \& Schechter 1974]{presschechter74}
   Press W. H., \& Schechter P., 1974, \apj, 187, 425

\bibitem[Pyne \& Birkinshaw 1993]{PyneBirkinshaw93}
   Pyne, T., Birkinshaw, M., 1993, \apj, 415, 459

\bibitem[Rees \& Sciama 1968]{ReesSciama68}
   Rees, M.J., \& Sciama, D.W., 1968. \nat, 217, 511

\bibitem[Rephaeli 1995]{Reph95} 
   Rephaeli, Y., 1995, \araa, 33, 541

\bibitem[Schneider, Ehlers \& Falco 1992]{Schneideret92}
   Schneider, P., Ehlers, J., \& Falco, E. E., 1992, ``Gravitational Lenses'', Springer: Berlin

\bibitem[Seljak 1996]{Seljak96}
   Seljak, U., 1996, \apj, 460, 549

\bibitem[Seljak 1997]{Seljak97}
   Seljak, U., 1997, preprint, astro-ph/9711124

\bibitem[Stompor \& Efstathiou 1998]{StomporEfstath98}
   Stompor, R., \& Efstathiou, G., 1998, preprint, astro-ph/9806294

\bibitem[Sunyaev \& Zel'dovich 1980]{sz80}  
   Sunyaev, R. A., \& Zel'dovich, Y. B., 1980, \araa, 18, 537

\bibitem[Suto et al. 1999]{sutoet99}
   Suto et al., 1999, Advances in Space Research, in press

\bibitem[Tegmark 1998]{Tegmark98} 
   Tegmark, M., 1998, \apj, 502, 1

\bibitem[Tegmark et al. 1998]{Tegmarket98} 
   Tegmark, M., Eisenstein, D. J., Hu, W., \& Kron, R. G., 1998, preprint, astro-ph/9805117

\bibitem[Toffolatti et al. 1999]{ToffZAB99}
   Toffolatti, L., De Zotti, G., Argueso, F., \& Burigana, C., 1999, preprint, astro-ph/9902343

\bibitem[Tuluie, Laguna \& Anninos 1996]{TuluiLagA96}
   Tuluie, R.,  Laguna, P., \& Anninos, P., 1996, \apj, 463, 15

\bibitem[Vishniac 1987]{Vish87}
   Vishniac, E. T., 1987, \apj, 322, 597 

\bibitem[Voit \& Donahue 1998]{VoitDonahue98}
   Voit, G. M., \& Donahue, M., 1998, preprint, astro-ph/9804306

\bibitem[White \& Srednicki 1995]{WhiteSrednicki95} 
   White, M., \& Srednicki, M., 1995, \apj, 443, 6

\bibitem[Winitzky 1998]{Winitzky98}
   Winitzky, S., 1998, preprint, astro-ph/9806105

\bibitem[Zaldarriaga \& Seljak 1998]{ZaldarriagaSeljak98}
    Zaldarriaga, M., \& Seljak, U., 1998, preprint, astro-ph/9803150

\bibitem[Zaldarriaga, Spergel \& Seljak 1997]{ZaldarrSperSel97}
    Zaldarriaga, M., Spergel, D. N., \& Seljak, U., 1997, \apj, 488, 1


\end{thebibliography}
\end{document}